%% file: main.tex
\documentclass[a4paper,fleqn, review]{cas-sc}
\usepackage{lineno}
\usepackage[numbers]{natbib}

\usepackage{amssymb}
\usepackage{caption}
\usepackage{subcaption}

\usepackage{cleveref}

\crefname{table}{Tab.}{Tabs.}
\crefname{figure}{Fig.}{Figs.}
\crefname{section}{Sec.}{Secs.}
\crefname{equation}{Eq.}{Eqs.}

\usepackage[ruled,vlined, norelsize]{algorithm2e}
\SetKwInput{KwInput}{Input}
\SetKwInput{KwOutput}{Output}

\newcommand{\red}[1]{\textcolor{red}{#1}}
\newcommand{\blue}[1]{\textcolor{blue}{#1}}
\newcommand{\green}[1]{\textcolor{teal}{#1}}

\def\tsc#1{\csdef{#1}{\textsc{\lowercase{#1}}\xspace}}
\tsc{WGM}
\tsc{QE}
\tsc{EP}
\tsc{PMS}
\tsc{BEC}
\tsc{DE}

\begin{document}
\let\WriteBookmarks\relax
\def\floatpagepagefraction{1}
\def\textpagefraction{.001}

\shorttitle{A Transparency Paradox? Impact of Explanation Specificity and Autonomous Vehicle Perceptual Inaccuracies}

\shortauthors{D. Omeiza et~al.}

\title [mode = title]{A Transparency Paradox? Investigating the Impact of Explanation Specificity and Autonomous Vehicle Perceptual Inaccuracies on Passengers}                      





%


\author[1]{Daniel Omeiza}[type=author,
                        orcid=0000-0002-1232-346X]

\cormark[1]
\cortext[cor1]{Corresponding author}
\ead{danielomeiza@robots.ox.ac.uk}


\credit{Conceptualisation of idea, writing and editing the original draft, software implementation, lab experiment, data collection, data analysis}

\affiliation[1]{organization={University of Oxford},
    city={Oxford},
    country={United Kingdom}}

\author[1,3]{Raunak Bhattacharyya}[style=english]
\ead{raunak@robots.ox.ac.uk}
\fnmark[1]

\credit{Writing and editing, review, data collection, lab experiment}

\author[1]{Marina Jirotka}[%
   ]
\ead{marina.jirotka@ox.ac.uk}

\credit{Conceptualisation, bi-weekly supervision and guidance, resources}

\author%
[1]
{Nick Hawes}
\ead{nickh@robots.ox.ac.uk}
\credit{Supervision, provision of resources}

\author[1,2]{Lars Kunze}[style=english]
\credit{Conceptualisation of idea, weekly supervision, and guidance, resources provision, review}

\affiliation[2]{organization={University of the West of England},
    city={Bristol},
    country={United Kingdom}}
    
\affiliation[3]{organization={Indian Institute of Technology},
    city={Delhi},
    country={India}}

\fntext[fn1]{This author was with the University of Oxford at the time this work was done. This author's affiliation is now the Indian Institute of Technology, Delhi}



\begin{abstract}
 Transparency in automated systems could be afforded through the provision of intelligible explanations. While transparency is desirable, might it lead to catastrophic outcomes (such as anxiety), that could outweigh its benefits? It's quite unclear how the specificity of explanations (level of transparency) influences recipients, especially in autonomous driving (AD). In this work, we examined the effects of transparency mediated through varying levels of explanation specificity in AD. We first extended a data-driven explainer model by adding a rule-based option for explanation generation in AD, and then conducted a within-subject lab study with 39 participants in an immersive driving simulator to study the effect of the resulting explanations. Specifically, our investigation focused on: (1) how different types of explanations (specific vs. abstract) affect passengers' perceived safety, anxiety, and willingness to take control of the vehicle when the vehicle perception system makes erroneous predictions; and (2) the relationship between passengers' behavioural cues and their feelings during the autonomous drives. Our findings showed that passengers felt safer with specific explanations when the vehicle's perception system had minimal errors, while abstract explanations that hid perception errors led to lower feelings of safety. Anxiety levels increased when specific explanations revealed perception system errors (high transparency). We found no significant link between passengers' visual patterns and their anxiety levels. Our study suggests that passengers prefer clear and specific explanations (high transparency) when they originate from autonomous vehicles (AVs) with optimal perceptual accuracy.


\end{abstract}




\maketitle
\input{introduction}

\section{Background}
\label{sec:rel_work}
This section provides a background on explanations for automated decisions and in autonomous driving from the literature.

\subsection{Explanations for Automated Decisions}
\label{sec:explanations}
The concept of explanations has been studied extensively by a multidisciplinary group of scholars, ranging from Philosophy to Psychology. Each adopts an idiosyncratic lens, characterising explanations in terms relevant to the goals of their respective disciplines. Therefore, to guide our study, we align with a definition proposed in an extensive survey of explanations in autonomous driving~\cite{omeiza2021explanations}: explanations are a piece of information presented in an intelligible way as a reason or part of a reason for an outcome, event or an effect in text, speech, or visual forms.

Recent efforts around explanations have been mainly channelled towards complex AI systems (explainable AI~\cite{gunning2017explainable, adadi2018peeking, dovsilovic2018explainable}) to understand and communicate the reasons for the systems' decisions.
Techniques developed for this purpose fall under different categories based on their mode of operation. Some are model-specific~\cite{chakraborti2020emerging} in that they investigate the underlying AI algorithm in detail to support debugging tasks~\cite{magnaguagno2017web}. These types of explainers are intrinsic and model specific meaning that they are inherently coupled with the underlying algorithm. Other explanation approaches are classified as model-agnostic as they can assess the properties of an output independent of the algorithms used to realise the output~\cite{ribeiro2016should,lundberg2017unified}. These explainers aim to help enhance user's knowledge of a system and potentially foster trust.

Explanations must possess certain properties to be effective to their intended recipients. Mittelstadt et al.~\cite{mittelstadt2019explaining} argued that the risk of conflicts in communicating explanations when the \textit{explainer} (explanation provider) and the \textit{explainee} (explanation recipient) have different motives may be mitigated through social, selective, and contrastive explanations. Social in the sense that the explanation process involves different parties and the explainer is able to model the expectations of the explainee. The explanation is selective if it can select explanations from several competing hypotheses. It is considered contrastive if it can differentiate the properties of two competing hypotheses. Kment~\cite{kment2006counterfactuals} further emphasised the value of counterfactual explanations in enhancing understanding. These explanations describe how changes in input can lead to a shift from one fact to a competing hypothesis (foil)~\cite{miller}. Explanations should be intelligible~\cite{omeiza2021towards} and strike a delicate balance between providing sufficient detail and respecting the cognitive capacity of the explainee. This balance presents a significant challenge, as accurately gauging an individual's real-time cognitive capacity remains difficult.

Different methodologies have been adopted in the XAI literature. Wang et al.~\cite{wang2019designing} categorised these research methodologies into three groups: First, the existence of \textit{unvalidated guidelines} for the design and evaluations of explanations was highlighted. The authors claimed that these kinds of guidelines are based on authors' experiences with no further substantial justification.
Second, researchers suggested (in~\cite{zhu2018explainable}) that understanding users' requirements is helpful in XAI research. It is on this premise that previous research on explanation design has been thought to be \textit{empirically derived}. This type of XAI research elicits explanation requirements from user surveys to determine the right explanation for a use case with explanation interfaces.
Third, some explanation design methods are derived from \textit{psychological constructs from formal theories} in the academic literature. Some of these methods (e.g., in \cite{hoffman2017explaining}) draw on theories from cognitive psychology to inform explanation design for explanation frameworks. Our work is heavily grounded on empirical studies.

\subsection{Explanations in Autonomous Driving}
\label{sec:rel_work_exp_ad}
Explanations have been found useful in enhancing user experience~\cite{schneider2021explain}, trust~\cite{koo2015did, ha2020effects}, and improved situational awareness~\cite{omeiza2021not,liu2021importance} in automated driving. Recent works have explored human factors in the application of explainable AI in AD. For instance, in~\cite{omeiza2021towards, omeiza2022spoken}, a socio-technical approach to explainability was proposed. An interpretable representation and algorithms for explanations based on a combination of actions, observations, and road rules were designed. Regarding explanations depths, the notion that explanations with higher levels of abstraction and correctness are superior has been argued in the literature~\cite{buijsman2022defining, guidotti2018survey}. Additionally, Ramon et al. (2021) argued that the specificity of explanations should be tailored to the application context, noting that low-level specificity is often preferred by individuals with a more deliberative cognitive style.

In this paper, the term \textit{explanation specificity}
is used to refer to two specificity levels of explanations, \textit{abstract} (low transparency) and \textit{specific} (high transparency). Explanations can be used to convey different information in AD, e.g., vehicle uncertainties and intentions, and communicated through different modalities. For example,
Kunze et al.~\cite{kunze2019automation} conveyed visual uncertainties with multiple levels to operators using heartbeat animation. This information helped operators calibrate their trust in automation and increased their situation awareness. Similarly, Kunze et al.~\cite{kunze2019conveying} used peripheral awareness display to communicate uncertainties to alleviate the workload on operators simultaneously observing the instrument cluster and focusing on the road. This uncertainty communication style decreased workload and improved takeover performance. In addition, the effects of augmented reality visualisation methods on trust, situation awareness, and cognitive load have been investigated in previous studies using semantic segmentation~\cite{colley2021effects}, scene detection and prediction~\cite{colley2022effects}, and pedestrian detection and prediction~\cite{colley2020effect}. These deep vision-based techniques applied to automated driving videos and rendered in augmented reality mode were a way of calling the attention of operators to risky traffic agents in order to enhance safety. While under-explored, auditory means of communicating explanations are important to calling in-vehicle participants' attention to critical situations in AD. We thus used an auditory communication style in this study to convey explanations to passengers.
Some existing works around human-machine interaction~\cite{liu2021importance} have leveraged theoretical models (e.g., mental and situational models~\cite{endsley2000situation}) to study explanations. We based our work on behavioural cues and subjective feedback from participants while drawing connections to such existing works.

\subsection{Research Questions}
From the preceding literature review, we find the need to gain a deeper understanding of the effects of transparency, brought about by natural language explanations 
 of varying specificity, especially under imperfect AV perception systems.
\label{sec:hypo}
\begin{enumerate}
    \item Given varying levels of perception system errors, how do natural language explanations influence passengers' perceived safety?
    \begin{itemize}
        \item \textbf{H1.1 - Perceived Safety.} \textit{Low transparency yields a higher perception of safety in an AV with perception system errors.} We hypothesise that passengers feel safer in a low transparency AV, despite receiving abstract explanations. While individuals often seek the truth, many prefer information that aligns with their expectations~\cite{hart2009feeling}. Consequently, specific explanations may reveal perception system errors that contradict passenger expectations. Additionally, research has shown that placebo explanations can have similar positive effects on people as real explanations~\cite{eiband2019impact}.
        
        \item \textbf{H1.2 - Feeling of Anxiety.} \textit{Passengers' feeling of anxiety increases with increasing perception system errors in a highly transparent AV.} We posit that there is a connection between perceived safety and the feeling of anxiety~\cite{davidson2016mediating, quansah2022perceived}. Therefore, explanations that frequently reference misclassified actors are likely to create a sense of insecurity, leading to increased anxiety.
    
        \item \textbf{H1.3 - Takeover Feeling.} \textit{In highly transparent AVs, passengers are more likely to develop the feeling to take over navigation control from the AV with higher perception system errors.} Although passengers are not able to take control in this study, we anticipated that they might nurse the thought to do so if they repeatedly received illogical explanations from the AV.
    \end{itemize}
    
    \item Do passengers' behavioural cues correlate with their feelings?
        
    \begin{itemize}
        \item \textbf{H2.1 - Visual Feedback} \textit{Visual feedback from participants correlates with their feeling of anxiety.} Individuals with the feeling of anxiety might be usually hyper-aroused and sensitive to environmental stimuli. They may have difficulties concentrating, performing tasks efficiently, and inhibiting unwanted thoughts and distractions ~\cite{hepsomali2017pupillometric, chen2014biased}. Participants' fixation points and saccades should correlate with anxiety.
    \end{itemize}
\end{enumerate}

\section{Passenger Study}
\label{sec:userstudy}
In this section, we describe the participants' demographic, experiment apparatus setup, experiment design, and the procedure of the experiment. The necessary approval to conduct the study was obtained from our University's Research Ethics Committee.

\subsection{Participants}
We conducted a power analysis to estimate the number of subjects required for the study. Afterward, calls for participants were placed on various online platforms, such as the callforparticipants platform, university mailing groups, university Slack channels, the research group website, and social media to recruit subjects. Upon screening, the final sample consisted of $N = 39$ participants (28 male, 11 female) ranging in age from 18 to 59 years.
The participants comprised students, university employees, and members of the callforparticipants platform. Although prior driving experiences were not required, 28 (71.79 \%) of the participants were licensed drivers.
Only 2 of the 39 participants (5.13\%) had experience with autonomous drives, however, in a research context. 6 (15.38\%) of the participants had used a virtual reality headset for a driving game or driving experiment in the past.

{\fontsize{8pt}{10pt}
\selectfont
\begin{table*}[]
\centering
\caption{\footnotesize Description of a subset of the events (5 out of 9) and corresponding explanations provided during the study. Observations and causal explanations are announced to passengers. AV's action (text in red), other agent's class \& action (text in blue), and the agent's location (text in green) are determined by the explainer algorithm described in Algorithm~\ref{alg:tb}.}

\begin{tabular}{@{}p{2.5cm} p{4.5cm} p{3.5cm} p{3.5cm}@{}}
\toprule
\textbf{Event} &
  \textbf{Description} &
  \textbf{Observation} &
  \textbf{Causal Explanation} \\ \midrule
FollowLeadingVehicle &
  AV follows a leading actor. At some point, the leading actor slows down and finally stops. The AV has to react accordingly to avoid a collision. &
  \blue{vehicle ahead} on \green{my lane}. &
  \red{Stopping} because \blue{cyclist stopped} on \green{my lane}. \\
VehicleTurning &
  AV takes a right or a left turn from an intersection where an actor suddenly drives into the way of the AV,  AV stops accordingly. After some time, the actor clears the road, AV continues driving. &
  \blue{motorbike crossing} \green{my lane}. &
  \red{Stopping} because \blue{motorbike is crossing} \green{my lane}. \\
LaneChangeObstacle &
  AV follows a leading actor, and at some point, the leading actor decelerates. The AV reacts accordingly by indicating and then changing lanes. &
  \blue{vehicle ahead} on \green{my lane}. &
  \red{Changing lane to the {[}right/left{]}} because \blue{vehicle stopped} on \green{my lane}. \\
StopSignalNoActor &
  No actor ahead of the AV at a signalised intersection with a red traffic signal. AV decelerates and stops. &
  \blue{red traffic light} ahead on \green{my lane}. &
  \red{Stopping} because \blue{traffic light is red} on \green{my lane}. \\
MovSignalNoActor &
  No actor ahead of the AV. AV starts moving from a stop state at a signalised junction or intersection. &
  None &
  \red{Moving} because \blue{traffic light is green} on \green{my lane}. \\
\bottomrule
\end{tabular}
\label{tab:scenarios}
\end{table*}
}

\subsection{Apparatus}
\subsubsection{Hardware}
The hardware setup is shown in \Cref{fig:setup}.
We conducted the experiment in a driving simulator that comprised a GTR arcade seat, Logitech G29 steering wheel with force feedback, turn signal paddles, brake and accelerator pedals, and an ultra-wide LG curved screen to display the experiment. A state-of-the-art virtual reality (VR) headset (with an immersive $360^{\circ}$ FoV and an eye tracker) was also used  to provide an immersive experience and high visual fidelity.
\begin{figure}
\centering
  \includegraphics[height=4.7cm]{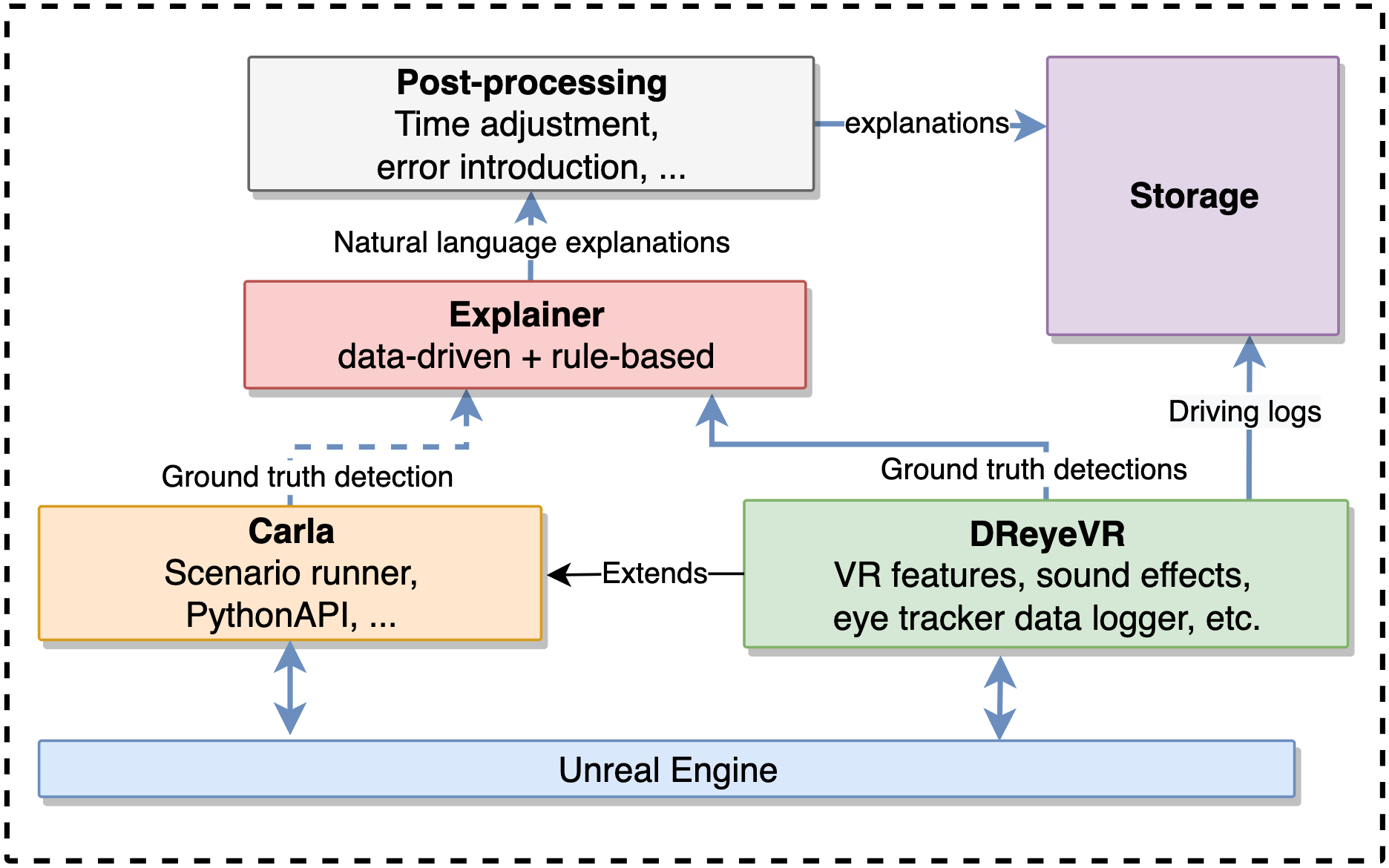}
  
  \caption{ \footnotesize High-level architecture of our simulation software. DReyeVR uses Unreal engine and extends  CARLA simulator, which also builds on Unreal engine. DReyeVR extends CARLA by adding VR functionalities, vehicular and ambience sounds, eye tracker data logging, and additional sensors, among others. Our explainer model, which is both rule-based and data-driven, receives ground truth data from  CARLA or DReyeVR and generates explanations for predicted actions. The post-processing script allowed us to modify the generated explanations as we desire.}~\label{fig:arch}
\end{figure}

\subsubsection{Driving Software}
Software architecture is illustrated in \Cref{fig: arch}.
We adapted the DReyeVR~\cite{silvera2022dreyevr}, an open-source VR-based
driving simulation platform for behavioural and interaction research involving human drivers. DReyeVR was built atop CARLA~\cite{dosovitskiy2017carla}, an open-source driving simulator for AD and Unreal Engine 4. DReyeVR provides a very realistic experience with naturalistic visuals (e.g., in-vehicle mirrors) and auditory (e.g. vehicular and ambient sounds) interfaces allowing for an ecologically valid setup. It also provides an experimental monitoring and logging system to record and replay scenarios, as well as a sign-based navigation system.

\subsubsection{Explainer Software}
 As shown in \Cref{fig:arch}, we developed an explainer system based on previous work~\cite{omeiza2022spoken}. This system utilises a tree-based model trained on an annotated AV driving dataset that we collected in a prior project. While the original algorithm in~~\cite{omeiza2022spoken} is primarily data-driven, we incorporated a rule-based technique to serve as a fallback when the data-driven method fails or makes an incorrect ego action prediction. The data-driven method employs a trained tree-based model to predict and generate explanations from detections obtained from CARLA, a driving simulator. In contrast, the rule-based approach relies on CARLA's ground truth data and follows predefined rules to determine which agents to reference in the explanations. By comparing predictions from the data-driven method with ground truth observations from CARLA, we can identify incorrect predictions. This enhanced explainer system, combining both data-driven and rule-based approaches, was used to generate preliminary explanations for our created scenarios. Wintersberger et al. (2020) suggested types of traffic elements to be included in visual explanations based on user preferences. Our proposed explainer, however, selects traffic elements deemed important (feature importance~\cite{anjomshoae2021context}) by the driving model for its decisions (see Algorithm~\ref{alg:tb}).

We performed post-processing operations on the generated explanations, including fine-tuning some of the content and adjusting timestamps to ensure the explanations were delivered at the appropriate moments.

\begin{algorithm}
\LinesNumbered
\DontPrintSemicolon
\KwInput{tree model $\mathcal{M}$ for ego's action prediction, input vector $\mathbf{X}$ describing ego's environment}
\KwOutput{intelligible auditory explanation}
    Select a representative tree $m \in \mathcal{M}$ from tree model $\mathcal{M}$.\\
    Predict action $y \in \mathcal{Y}$ given $\mathbf{X}$.\\
    Compare prediction $y$ with CARLA ground truth $y_{GT}$.\\
    \uIf{$y = y_{GT}$}{
        Trace the decision path $\mathcal{P}_y$ for the prediction $y$ in tree $m$.\\
        Compute the importance score $I(\mathbf{X}_i)$ of the attributes $\mathbf{X}_i$ in each node $i$ along the decision path $\mathcal{P}_y$.\\
        Select attributes $\mathbf{X}_i$ with importance scores $I(\mathbf{X}_i) \geq k$, where $k$ is a predefined threshold.\\
        Merge the conditions/inequalities in the selected attributes $\mathbf{X}_i$.\\
        Translate merged attributes $\mathbf{X}_i$ to natural language following the template in \Cref{tab:scenarios}.\\
    }
    \Else{
        Use CARLA ground truth information $y_{GT}$ and predefined rules to generate explanation following the template in \Cref{tab:scenarios}.\\
    }
    \caption{Intelligible Explanation Generation}
    \label{alg:tb}
\end{algorithm}

\section{Experiment Design}
Before the start of the trials, participants were asked to manually drive a vehicle for about two minutes in CARLA Town03---a complex town, with a 5-lane junction, a roundabout, unevenness, and a tunnel. 30 vehicles and 10 pedestrians were spawned in this town. This preliminary drive aimed to familiarise participants with the driving simulation environment and to allow them to experience manual driving within the simulation. 

We employed a within-subject design due to the limited sample size, which was insufficient for a between-subject study. Additionally, this design helped mitigate the potential co-founding factor of between-individual differences in a between-subject design.

\subsection{Independent Variable}
\begin{table}[!htp]\centering
\caption{Description of Scenarios: The Independent Variables}\label{tab:variables}
\scriptsize
\begin{tabular}{lrrrrrr}\toprule
\textbf{Scenario} &\textbf{Specificity/Transparency} &\textbf{Perception Errors (\%)} &\textbf{Scenario Length} &\textbf{Examples of Events} &\textbf{Environment} \\\midrule
\textbf{Abstract} &Low &0 & ${\!\sim \!4}$ minutes &All events from Table 1 &CARLA Town10HD \\
\textbf{Specific (5)} &High &5 &${\!\sim \!4}$ minutes &All events from Table 1 &CARLA Town10HD \\
\textbf{Specific (50)} &High &50 &${\!\sim \!4}$ minutes &All events from Table 1 &CARLA Town10HD \\
\bottomrule
\end{tabular}
\end{table}

Combinations of transparency level (low and high) and AV perception errors (low and high) were done to obtain the independent variable \textit{Scenarios}. The first scenario (\textit{Abstract} scenario) comprises abstract explanations indicating low transparency and an undefined amount of perception system errors. The second scenario (\textit{Specific(5)} scenario) comprises specific explanations indicating high transparency and 5\% amount of perception system errors indicating low error degree. The third scenario (\textit{Specific(50)} scenario) comprises specific explanations indicating high transparency and 50\% amount of perception system errors indicating high error degree.
\begin{figure}[h!]
         \centering
         \includegraphics[height=5cm]{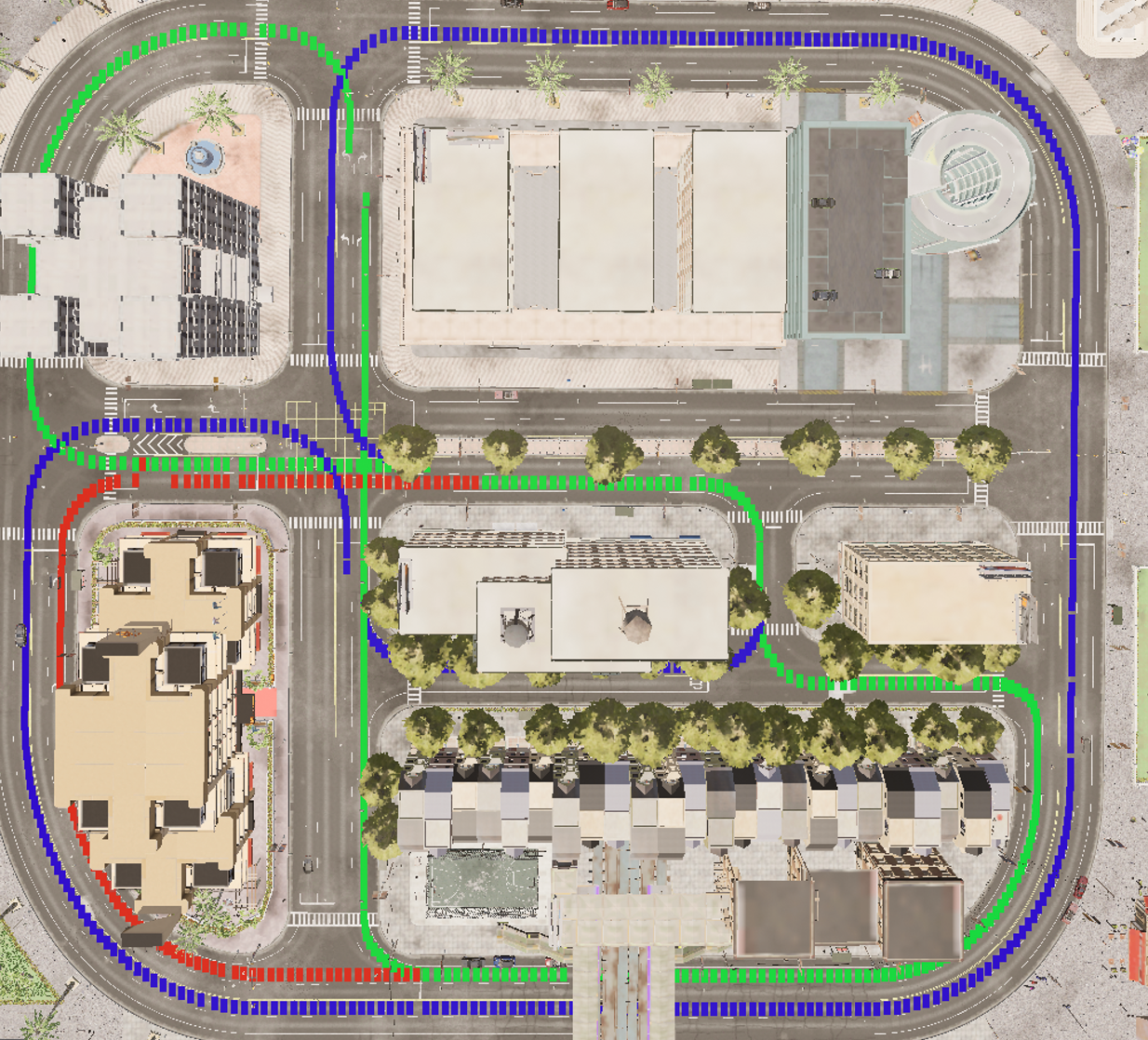}
         \caption{Scenario routes. \textbf{Red}: Abstract, \textbf{Green}: Specific(5), \textbf{Blue}: Specific(50). Each route is a loop and overlaps with others at some points.}
         \label{fig:map}
     \end{figure}
The driving events that made up the different scenarios were carefully designed to include different driving conditions that are obtainable in the real world (See Table~\ref{tab:scenarios}). Note that scenarios 
The scenario routes are shown in Figure~\ref{fig:map}.
\paragraph{\textbf{i. Abstract}} A scenario in CARLA Town10HD, which is about 4 minutes long (330 secs). Town10HD is an urban city environment with different infrastructures, such as an avenue or promenade, and realistic textures. Driving conditions are a combination of the events in Table~\ref{tab:scenarios}. The perception system in this scenario might contain some errors, but the explanations provided in this scenario were post-processed to always provide surface information which is vague enough to conceal perception errors. The rules governing explanations for this scenario were:
\begin{itemize}
\item all traffic lights are referred to as `traffic sign' without specifying the state (e.g., red, green, amber, off) of the traffic light;
\item pedestrians are referred to as `road users';
\item all non-human moving actors are referred to as `vehicle'. This includes cycles, motorbikes, cars, etc.
\end{itemize}
An example explanation is `stopping because of the traffic sign on my lane'. This obfuscates the type and colour of the traffic sign.

\paragraph{\textbf{ii. Specific(5)}} A scenario in CARLA Town10HD, which was about 4 minutes in length (256 seconds). Driving conditions in this scenario were a combination of the events in Table~\ref{tab:scenarios}. The explanations generated in this scenario were specific and detailed, exposing all errors. The perception system of the AV in this scenario was about 5\% inaccurate. This error value was estimated following the dynamic traffic agent classification model and confusion matrix provided by ~\cite{bin2021double} and the traffic light classification model and confusion matrix by~\cite{michael2015extending}. We were only interested in the confusion matrices (and not the models). The confusion matrices helped us to systematically introduce the 5\% perception system errors during the post-processing stage of the explanations. In this scenario, the 5\% error resulted in one explanation (1 out of the 22) being erroneous as the explanation exposed the misclassification errors from the perception system. 
An example of an erroneous explanation is: `van ahead on my lane'. Here, a car was misclassified as a van.

\paragraph{\textbf{iii. Specific(50)}} A scenario in CARLA Town10HD, which was 4 minutes in length (274 seconds). Driving conditions were a combination of the events in Table~\ref{tab:scenarios}. The explanations generated in this scenario were as fine-grained/specific and detailed as those in the \textit{Specific(5)} scenario. The perception system error of the AV in scenario \textit{Specific(5)} was significantly noised to reach a reduced accuracy of 50\%. We assumed that this reduction in accuracy might be sufficient to influence peoples' behaviour.
Therefore, half of the explanations in this scenario (12 out of 24) reflected misclassification of actors or actor states. An example of an erroneous explanation is `moving because traffic light is switched off on my lane'. In this case, the perception system failed to identify a green light accurately.

\begin{figure*}
     \centering
     \begin{subfigure}[b]{0.3\textwidth}
         \centering
         \includegraphics[width=4.6cm]{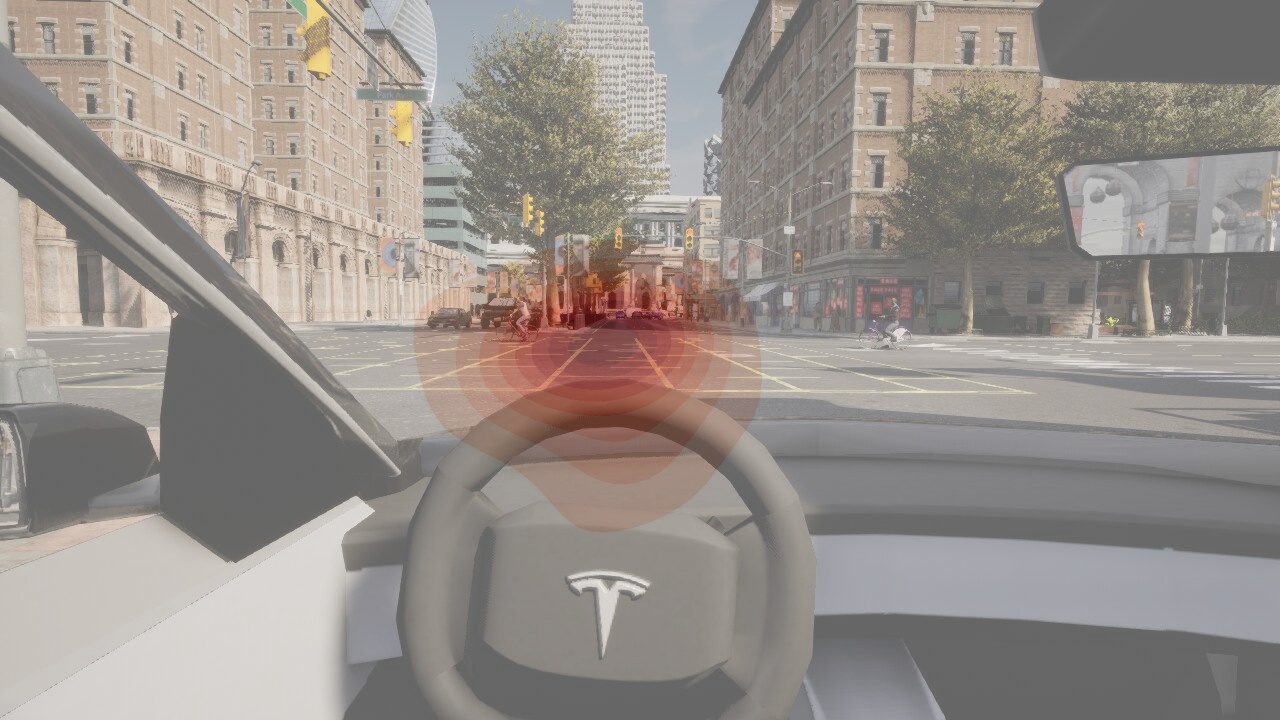}
         \caption{Abstract - \textit{Observation:} \textbf{`vehicle is crossing on my lane'\\}}
         \label{fig:vague}
     \end{subfigure}
     \hfill
     \begin{subfigure}[b]{0.3\textwidth}
         \centering
         \includegraphics[width=4.6cm]{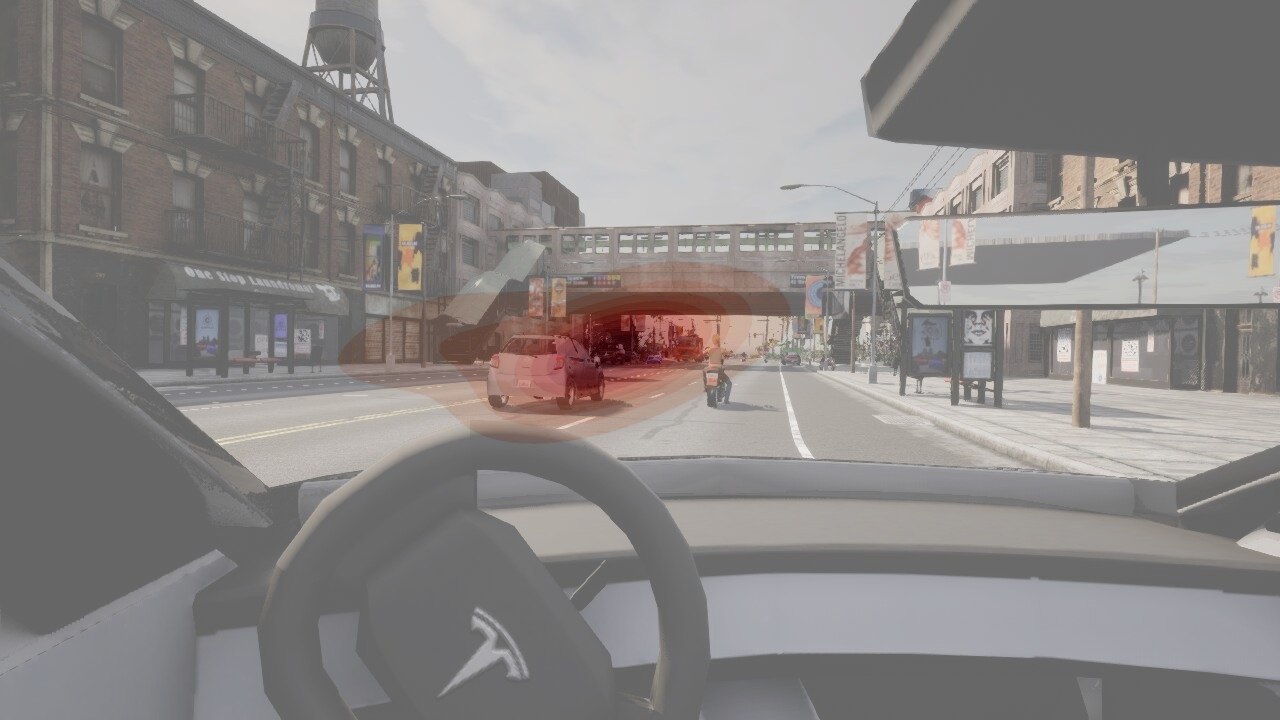}
         \caption{Specific(5) - \textit{Observation:} \textbf{ `motor bike ahead on my lane'\\}}
         \label{fig:spec5}
     \end{subfigure}
     \hfill
     \begin{subfigure}[b]{0.3\textwidth}
         \centering
         \includegraphics[width=4.6cm]{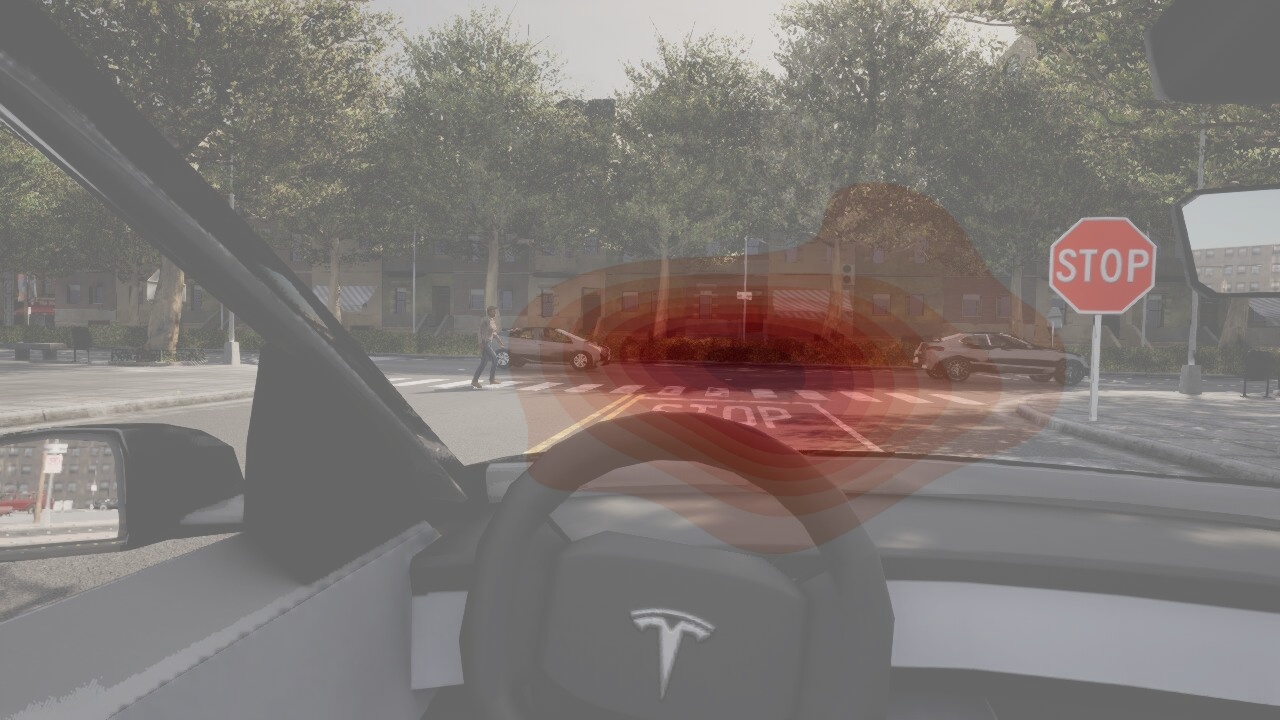}
         \caption{Specific(50) - \textit{Causal Explanation:} \textbf{`stopping because cyclist is crossing my lane.'}}
         \label{fig:spec50}
     \end{subfigure}
\caption[Sample screenshots and the generated explanations ]{Sample screenshots and the generated explanations (including observations announcement and causal explanations) from the three driving scenarios. Heatmaps of gaze points from all the participants are plotted over the images, indicating areas of interest. In the \textit{Abstract} scenario (Figure~\ref{fig:vague}), all movable/dynamic non-human actors are referred to as `Vehicle'. Thus, a cyclist was referred to as a vehicle. Figure~\ref{fig:spec5} depicts a scene from the \textit{Specific(5)} scenario in which the AV's perception system accurately identified and classified a motorbike and provided a fine-grained explanation for this. In the \textit{Specific(50)} scene (Figure~\ref{fig:spec50}), the AV's perception system misclassified a pedestrian as a cyclist. The fine-grained/specific explanation provided exposed this error.
}~\label{fig:sample}
\end{figure*}

Note that all three scenarios were designed so that the AV perception errors were insignificant to the AV's navigation actions. Hence, the AV respected all road rules and avoided collisions. This was important as the state-of-the-art AVs would likely not make obvious navigation errors. Moreover, we were interested in the effects of the awareness of inconsequential perceptual errors in AVs. Hence, it was necessary to introduce artificial errors of varying degrees (low and high). The non-influence of AV perception errors on navigation control also helped to avoid the confounding factors of route navigation problems. Further, we counterbalanced the routes across scenarios. That is, the AV's route was different in each scenario. This design decision was made to reduce carry-over effects on the participants. With this setup, the scenarios were still comparable as they were all within the same town, and the routes shared similar features. Each scenario also had a balanced combination of the events listed in Table~\ref{tab:scenarios}. In all the scenarios, the AV maintained a speed below $30mph$, the recommended speed limit in urban areas in the UK. See Figure~\ref{fig:sample} for sample scenes from each scenario and their corresponding explanations.

\subsection{Dependent Variables}
There were six dependent variables: \textit{Perceived Safety, Feeling of Anxiety, Takeover Feeling, Fixation Divergence, Saccade Difference,} and \textit{Button Presses}.
These variables were categorised into two (psychological factors and behavioural cues) for easy analysis and reporting.

\paragraph{\textbf{Psychological Factors}}
These factors include \textit{Perceived Safety, Feeling of Anxiety}, and \textit{Takeover Feeling}. They were mainly measured using items from the Autonomous Vehicle Acceptance Model Questionnaire (AVAM)~\cite{hewitt2019assessing}. AVAM is a user acceptance model for autonomous vehicles, adapted from existing user acceptance models for generic technologies. It comprises a 26-item questionnaire on a 7-point Likert scale, developed after a survey conducted to evaluate six different autonomy scenarios. 

Items 24---26 were used to assess the \textit{Perceived Safety} factor, while items 19---21 were used to assess the \textit{Feeling of Anxiety} factor. Similar to~\cite{schneider2021explain}, we introduced a new item to assess participants' feelings to takeover navigation control from the AV during the ride (\textit{Takeover Feeling}). Specifically, participants were asked to rate the statement `During the ride, I had the feeling to take over control from the vehicle' on a 7-point Likert scale. Actual navigation takeover by participants was not permitted because we wanted to be able to control the entire experiment and have all participants experience the same scenarios. Moreover, we were dealing with L4 automation. Though participants were not expected to drive or take over control, they might have nursed the thought to do so. This is what the \textit{Takeover Feeling} variable measures. 

We added a free-response question related to explanations with the aim of obtaining qualitative data for triangulating quantitative results. Participants were asked the following question: `What is your thought on the explanations provided by the vehicle, e.g., made you less/more anxious, safe, feeling to take over control?'. We refer to the resulting questionnaire as the APT Questionnaire (i.e., A-Anxiety, P-Perceived Safety, T-Takeover Feeling).

\paragraph{\textbf{Behavioural Cues}}
We also used \textit{Button Presses, Fixation Divergence}, and \textit{Saccade Difference} as additional metrics. \textit{Button Presses} were used to express unsafe, anxious or confused feelings.

\textit{Fixation Divergence} is the Euclidean distance between mean participants' fixation points and reference fixation points. This provides information to draw inferences about participants' distractions. 

For \textit{Saccade Difference}, we estimated participants' saccade velocity over time following the method in~\cite{gibaldi2021saccade} and found the difference from a reference saccade velocities. Saccade is the rapid movement of the eye between fixation points. Saccade velocity is the speed of such movements.
The fixation and saccade reference points (or ground truths) were the fixation and saccade records obtained from the researcher, who also participated in the study.


\begin{figure*}
\centering
  \includegraphics[height=3.5cm]{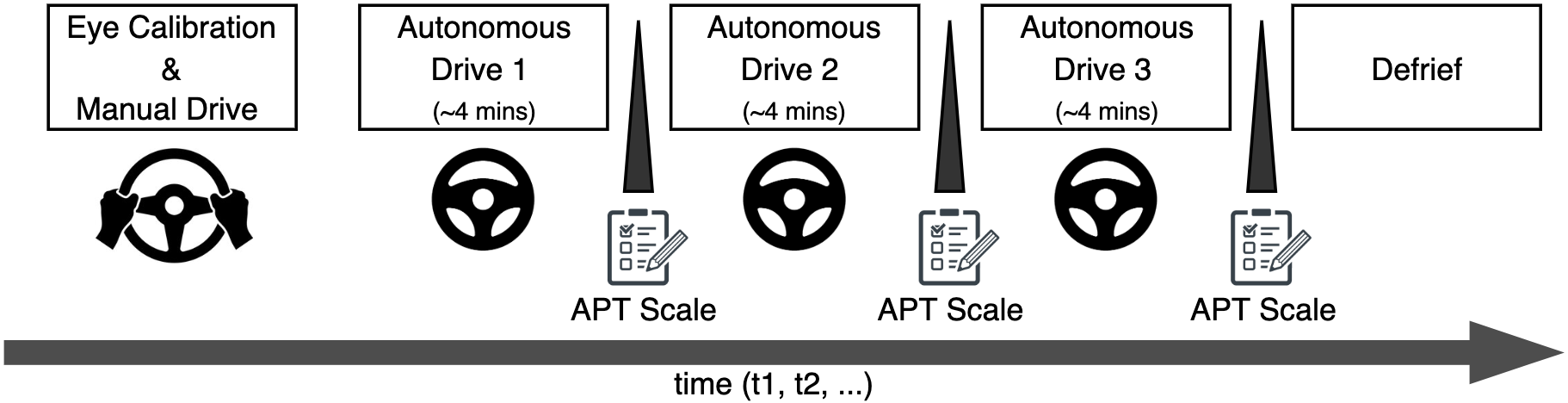}
  \caption[Study procedure]{Study procedure. Eye calibration was done with the VR headset; participants drove for two minutes, participants experienced each of the ~4 mins scenarios in counterbalanced order and completed the Feeling of Anxiety, Perceived Safety, and Takeover Feeling Questionnaire (APT Scale) in between each scenario. Participants were debriefed.}~\label{fig:flow}
\end{figure*}

\subsection{Procedure}
The procedure of the experiment is illustrated in \Cref{fig:flow}. After all preliminary form completions and briefings, we introduced the physical driving simulator and explained the subsequent steps, which involved a pre-experiment manual driving session in VR mode lasting for 2 minutes. Participants were informed that the purpose of this pre-experiment exercise was to familiarise them with the simulation environment and to identify individuals prone to motion sickness, who would then be excluded from the main experiment.

Upon completion of the manual driving exercise, the researcher removed the VR headset from the participant and explained the aim and procedure of the main experiment. The instructions included the following statements: `you would experience 3 autonomous rides by different vehicles, [...] and after each ride, you would complete a short survey. The vehicle drives along a predefined path for about 4 minutes and provides explanations for its planned driving decisions and announces relevant objects in its environment. [...]. The vehicle tells you its next direction at a junction or an intersection using its right or left red light indicators on its dashboard accordingly.[...] Simply click any of these buttons if the decision or the explanation of the vehicle makes you feel confused, anxious or unsafe [...]'. The researcher then placed the VR headset back on the participant and initiated the scenarios. Complete counterbalancing was applied to the scenario treatments, resulting in six different orders of scenarios. Each participant experienced the scenarios in one of these six orders, with approximately six participants per order.

Participants were encouraged to rest briefly after each driving experience, with the VR headset removed. A short debriefing session followed the study, and participants were given a £10 Amazon gift card. The entire experiment lasted approximately 50 minutes.

The researcher also participated in the experiment, experiencing all three scenarios. Throughout, the researcher focused on the lane ahead and the actors referenced by the explanations. Neither erroneous nor abstract explanations influenced the researcher's focus, as the researcher remained attentive to the lane and the actors or obstacles impacting the AV's actions, regardless of the explanations. This consistency was due to the researcher's familiarity with all scenarios. The data from the researcher served as a reference or ground truth. Notably, the researcher’s eye movements matched normal human saccadic velocity, which reaches 300---400\textdegree /seconds~\cite{raab1985normal,wilson1993saccadic}.

\section{Quantitative Results}
\subsection{Psychological Factors Analysis}
\begin{figure}[h!]
\centering
  \includegraphics[height=8.5cm]{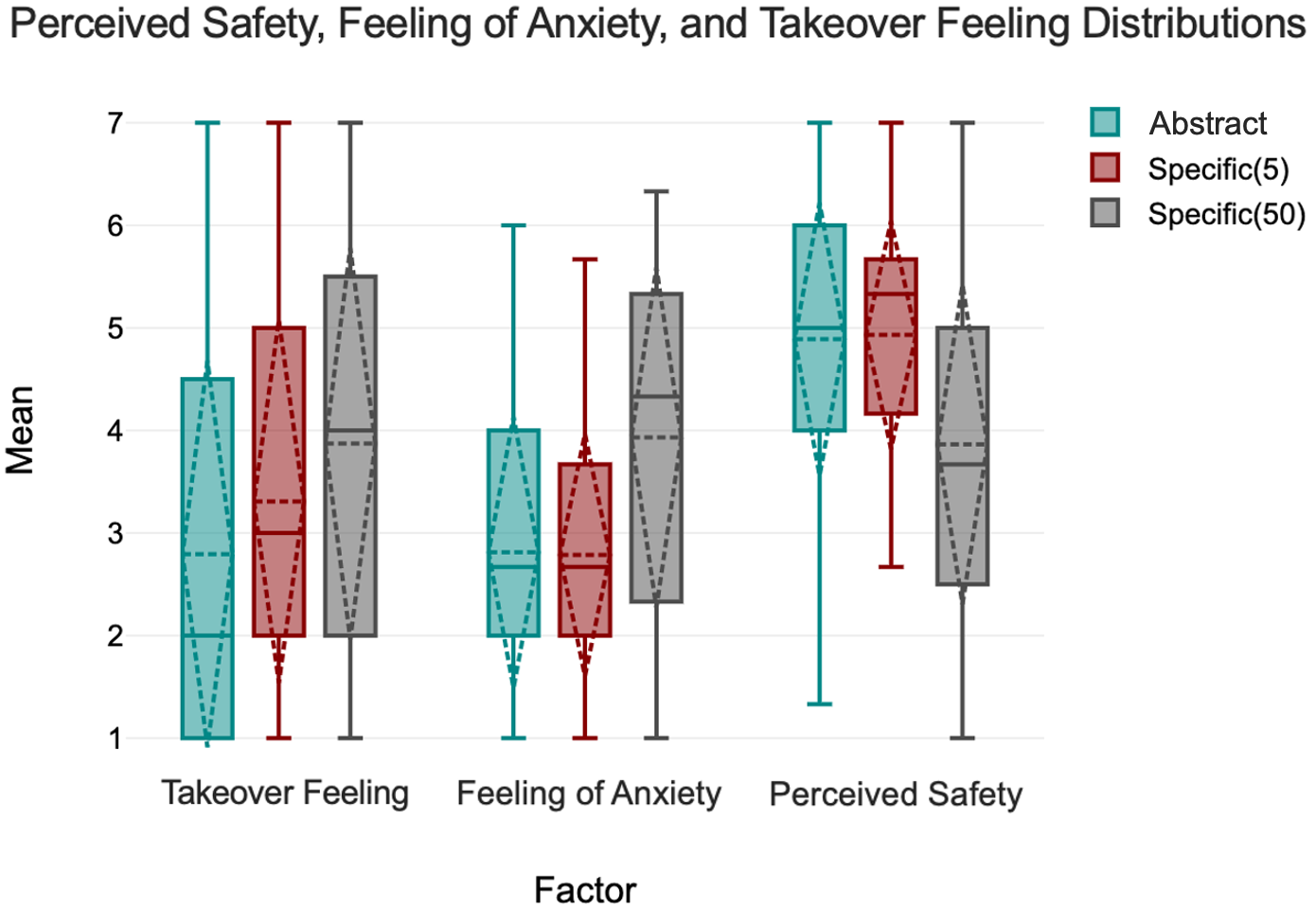}
  \caption[Perceived safety, feeling of anxiety, and takeover feeling distribution.]{Perceived safety, feeling of anxiety, and takeover feeling distribution. Perceived safety is highest in the Specific(5) scenario, the feeling of anxiety is highest in the Specific(50), and takeover feeling is lowest in the Abstract scenario.}~\label{fig:factors}
\end{figure}

To test our hypotheses listed in Section~\ref{sec:hypo},  we analyzed data from the three APT questionnaires. We created a latent variable (\textit{Feeling of Anxiety}) by averaging responses to AVAM Items 19---21, and another latent variable (\textit{Perceived Safety}) by averaging responses to AVAM Items 24---26. We calculated Cronbach's Alpha ($\alpha$) for the independent variables that formed the latent dependent variables to ensure they had adequate internal consistency. Results with an adjusted p-value less than 0.05 ($p < .05$) were considered significant. P-values were adjusted using Bonferroni corrections, where the calculated p-values were multiplied by the number of scenarios, to reduce the likelihood of Type I errors (false positives). Normality tests, including the Kolmogorov-Smirnov, Shapiro-Wilk, and Anderson-Darling tests, indicated a violation of normality in the \textit{Feeling of Anxiety}, \textit{Perceived Safety}, and \textit{Takeover Feeling} factors. Hence, we performed a Friedman test for these dependent variables,
see Table~\ref{tab:results} and Figure~\ref{fig:factors}.

\begin{table*}[]
{
\fontsize{9.5pt}{10.5pt}
\selectfont
\centering
\caption{Descriptive statistics from APT questionnaire analysis.}
\begin{tabular}{p{1.5cm} p{0.6cm} p{0.5cm}cccccccccl}\toprule
& \multicolumn{3}{c}{\parbox{3cm}{\textbf{Perceived Safety} \\Cronbach $\alpha: 0.87$, \\$H(2) = 8.17, \boldsymbol{p=.017}$}} & \multicolumn{3}{c}{\parbox{3cm}{\textbf{Feeling of Anxiety} \\Cronbach $\alpha: 0.86$\\ $H(2) = 13.32, \boldsymbol{p=.001}$}} & \multicolumn{3}{c}{\parbox{3cm}{\textbf{Takeover Feeling} \\ $H(2) = 6.27, \boldsymbol{p=.044}$}}
\\\cmidrule(lr){2-4}\cmidrule(lr){5-7} \cmidrule(lr){8-10}
& Mean  & SD & Mean Rank & Mean  & SD  & Mean Rank & Mean & SD  & Mean Rank \\\midrule
Abstract   & 4.89 & 1.35 & 2.15 & 2.81 & 1.34 & 1.72 & 2.79 & 1.91 & 1.68 \\
Specific(5) & \textbf{4.93} & 1.13 & \textbf{2.22} & 2.79 & 
1.2 & 1.81 & 3.31 & 1.79 & 2.10 \\
Specific(50) & 3.86 & 1.58 & 1.63 & \textbf{3.93} &
1.68 & \textbf{2.47} & \textbf{3.87} & 1.94 & \textbf{2.22} \\\bottomrule
\end{tabular}
\label{tab:results}
}
\end{table*}

\paragraph{\textbf{H1.1 - Perceived Safety}}
\textit{Low transparency yields a higher perception of safety in an AV with perception system errors.}

A Friedman test was conducted. No significant difference was found in the scenario pair: \textit{Abstract} — \textit{Specific(5)}, and the pair: \textit{Abstract} —
 \textit{Specific(50)}. In fact, the \textit{perceived safety} mean rank in the \textit{Specific(5)} scenario (2.22) was higher than that in the \textit{Abstract} scenario (2.15), see Table~\ref{tab:results}. Therefore, there was no sufficient evidence in support of hypothesis H1.1.

\paragraph{\textbf{H1.2 - Feeling of Anxiety}} \textit{Passengers' feeling of anxiety increases with increasing perception system errors in a highly transparent AV.}
A Friedman test indicated a significant difference in the \textit{Feeling of Anxiety} across scenarios, $H(2) = 13.32, p=.001$.
The pairwise scenario comparisons of \textit{Abstract} - \textit{Specific(50)} and \textit{Specific(5)} - \textit{Specific(50)} resulted in an adjusted p-value of $.003$ and $.01$ respectively (see Table~\ref{tab:results}).
Hence, there is strong evidence in support of hypothesis H1.2.

\paragraph{\textbf{H1.3 - Takeover Feeling}}
\textit{In highly transparent AVs, passengers are more likely to develop the feeling to take over navigation control from the AV with higher perception system errors. }
A Friedman test showed a significant difference in \textit{Takeover Feeling} across scenarios, $H(2) = 6.27, p=.044$.
While the pairwise scenario comparison of \textit{Abstract} - \textit{Specific(50)} resulted in an adjusted p-value of $.017$, the pairwise comparison of \textit{Specific(5)} - \textit{Specific(50)} resulted in an adjusted p-value of $0.61$.
Hence, there is no significant difference in 
 \textit{Takeover Feeling} between \textit{Specific(5)} and \textit{Specific(50)} scenarios, and therefore, no evidence in support of hypothesis H1.3 (see Table~\ref{tab:results}).

\begin{figure*}[h!]
\centering
  \includegraphics[width=\linewidth]{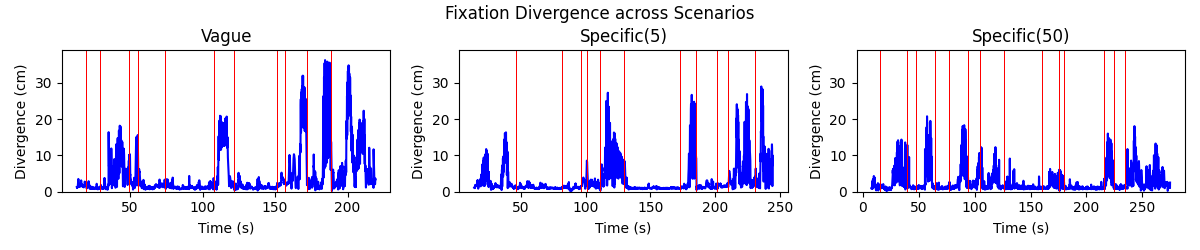}
  \caption[Fixation divergence across scenarios]{Fixation divergence across scenarios. While \textit{Specific(5)} had the highest mean fixation divergence, \textit{Specific(50)} had more frequent high fixation divergences. Red vertical bars represent the positions in time where causal explanations were provided.}~\label{fig:fixation}
\end{figure*}

\begin{figure*}
\centering
  \includegraphics[width=\linewidth]{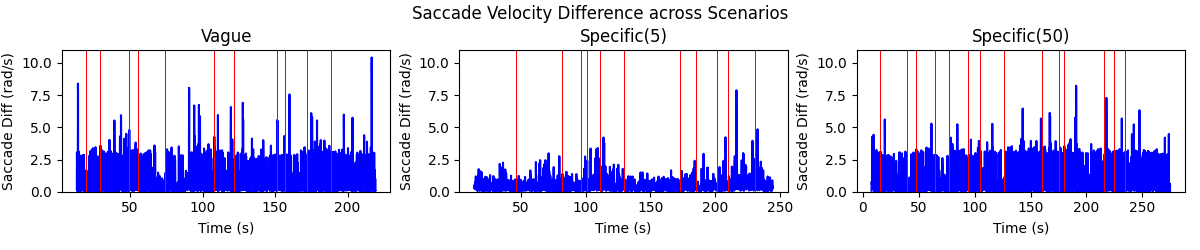}
  \caption[Saccade velocity difference across scenarios]{Saccade velocity difference across scenarios. \textit{Specific(5)} had the lowest mean saccade velocity difference while the \textit{Abstract} scenario had the highest. Red vertical bars represent the positions in time where causal explanations were provided.}~\label{fig:divergence}
\end{figure*}

\subsection{Behavioural Cues Analysis}
\paragraph{\textbf{H2.1 - Visual Responses}}
\textit{Visual feedback from passengers correlates with passengers' anxiety.}
At this stage, we utilised the reference data from the researcher. We estimated the Euclidean distances between participants' fixation points and the reference fixation points over time for each participant.

Results from Spearman correlation showed that there was no significant association between the \textit{Feeling of Anxiety} and \textit{Fixation Divergence}, $r(115) = -0.07, p = .442$. See the fixation divergence plot in Figure~\ref{fig:fixation}. Results from Spearman correlation showed that there was no significant association between the \textit{Feeling of Anxiety} and \textit{saccade difference}, $r(115) = 0.1, p = .281$. However, there was a significant association between \textit{perceived safety} and \textit{saccade difference}, $r(115) = -0.25, p = .007.$, indicating a weak negative correlation between \textit{perceived safety} and \textit{saccade difference}. 
Hypothesis H2.1, therefore, has no sufficient support. See the saccade difference plot in Figure~\ref{fig:divergence}.

In addition to correlation, we checked for significant differences. There was a significant difference in \textit{Fixation Divergence} between \textit{Abstract} and \textit{Specific(5)} with an adjusted p-value of $.028$, and between \textit{Specific(5)} and \textit{Specific(50)} with an adjusted p-value $< .001$. 
See Table~\ref{tab:metric2} for descriptive statistics.
Also, there was a significant difference between \textit{Abstract} and \textit{Specific(5)} with respect to \textit{Saccade Difference} (adjusted p-value of $<.001$). See Figure~\ref{fig:sample} for sample scenes from each scenario with the generated explanations. All the participants' gaze points are plotted as heatmaps over the screenshots.

\begin{table*}[]
{
\fontsize{9.5pt}{10.5pt}
\selectfont
\centering
\caption{Descriptive statistics from the haptic and Visual responses.}
\begin{tabular}{p{1.5cm} p{0.6cm} p{0.5cm}cccccccccl}\toprule
& \multicolumn{3}{c}{\parbox{3cm}{\textbf{ButtonPress} \\
$H(2) = 15.44, \boldsymbol{p<.001}$}} & \multicolumn{3}{c}{\parbox{3cm}{\textbf{Fixation Divergence} \\
$H(2) = 20.67, \boldsymbol{p<.001}$}} & \multicolumn{3}{c}{\parbox{3cm}{\textbf{Saccade Difference} \\ $H(2) = 15.35, \boldsymbol{p<.001}$}}
\\\cmidrule(lr){2-4}\cmidrule(lr){5-7} \cmidrule(lr){8-10}
& Mean  & SD & Mean Rank & Mean  & SD  & Mean Rank & Mean & SD  & Mean Rank  \\\midrule
Abstract   & 2.26 & 3.35 & 1.72 & 4.37 & 1.84 & 1.95 & \textbf{1.25} & 0.43 & \textbf{2.42} \\
\textit{Specific(5)} & 1.64 & 1.63 & 1.77 & \textbf{7.66} & 5.21 & \textbf{2.54} & 1.06 & 0.42 & 1.54 \\
Specific(50) & \textbf{4.9} & 4.33 & \textbf{2.51} & 3.2 & 1.23 & 1.51 & 1.17 & 0.45 & 2.04 \\\bottomrule
\end{tabular}
\label{tab:metric2}
}
\end{table*}

\subsubsection{ \textbf{Haptic Response}}
Participants were asked to press a button on the Logitech wheel when they felt confused, anxious, or unsafe by the explanations or the decision of the AV during the ride. Spearman rank correlation was used as a measure to investigate monotonic associations. 
\begin{figure}[h!]
\centering
  \includegraphics[height=8.5cm]{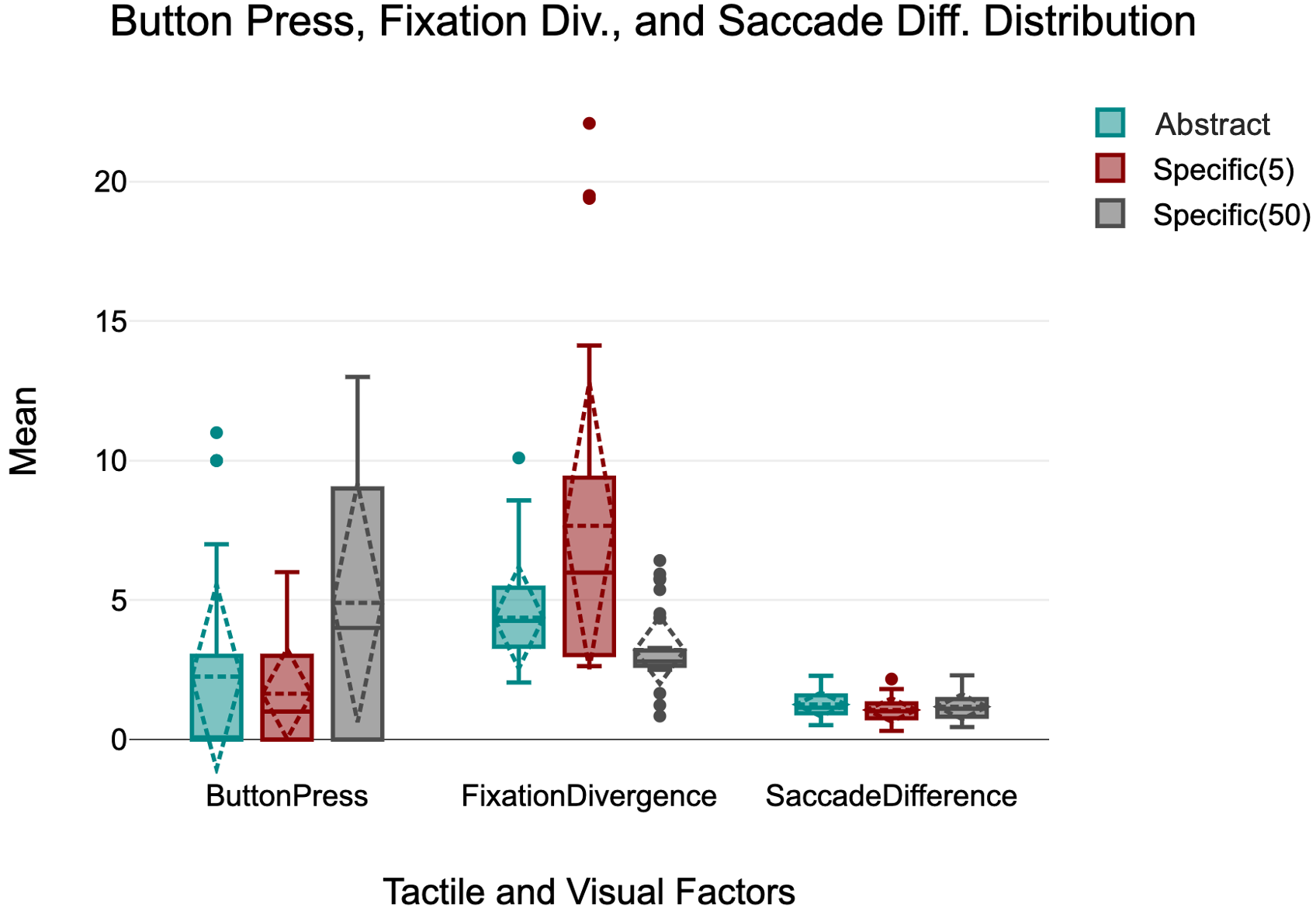}
  \caption[Button presses, fixation divergence and saccade difference distribution.]{Button presses, fixation divergence and saccade difference distribution. Button presses are fewer in the \textit{Specific(5)} scenario. Fixation divergence is highest in the \textit{Specific(5)} scenario, and saccade difference is lowest in the \textit{Specific(5)}. }~\label{fig:behaviour}
\end{figure}
There was a weak negative correlation between the variables \textit{Perceived Safety} and \textit{ButtonPress} ($r(115) = -0.31, p = .001$), a weak positive correlation between the \textit{Feeling of Anxiety} and \textit{ButtonPress} ($r(115) = 0.31, p = .001$), and insignificant correlation between the \textit{Feelings to Takeover} and \textit{ButtonPress} ($r(115) = 0.15, p = .099$).

We also checked for statistically significant differences in \textit{Button Presses} across scenarios. There was a significant difference in  \textit{ButtonPresses}, $H(2) = 15.44, p < .001$. This was specifically in the pairs: \textit{Abstract} - \textit{Specific(50)} with adjusted p-value $.002$, and \textit{Specific(5)} - \textit{Specific(50)} with adjusted p-value $.005$. See Figure~\ref{fig:behaviour} for behavioural cues results.

\section{Qualitative Results: Themes and Reflections}
\label{sec:reflection}

\begin{table*}[h!]
{\fontsize{9.5pt}{10.5pt}
\selectfont
\centering
\caption[Themes derived from the thematic analysis of the qualitative data from participants]{Themes derived from the thematic analysis of the qualitative data from participants. Freq. = Frequency of occurrence, SP = Scenario Percentage}
\begin{tabular}{l|p{3.2cm}lccccccccl}\toprule
& & \multicolumn{2}{c}{\parbox{2cm}{\textbf{Abstract} }} & \multicolumn{2}{c}{\parbox{2cm}{\textbf{Specific(5)}}} &
\multicolumn{2}{c}{\parbox{2cm}{\textbf{Specific(50)} }}
\\\cmidrule(lr){3-4} \cmidrule(lr){5-6} \cmidrule(lr){7-8}
Category & Theme  & Freq. & SP (\%) & Freq.  & SP (\%)  & Freq. & SP (\%) \\\midrule
Feelings   &
Anxious              & 2 & 5  & 2  & 5  & 8  & \textbf{21} \\
&Less Anxious         & 5 & \textbf{13} & 5  & \textbf{13} & 1  & 3  \\ 
&Safe                 & 9 & 23 & 12 & \textbf{31} & 7  & 18 \\ 
&Unsafe               & 0 & 0  & 1  & \textbf{3}  & 1  & \textbf{3}  \\ 
&Takeover             & 2 & 5  & 2  & 5  & 7  & \textbf{18} \\ 
&Confident            & 2 & 5  & 5  & \textbf{13} & 3  & 8  \\ 
&Trust                & 2 & \textbf{5}  & 1  & 3  & 2  & \textbf{5}  \\ 
&Distrust             & 1 & 3  & 0  & 0  & 6  & \textbf{15} \\ 
&Reassuring           & 5 & \textbf{13} & 2  & 5  & 0  & 0  \\ 
&Uncomfortable        & 2 & \textbf{5}  & 1  & 3  & 0  & 0  \\
\midrule
   
Explanations & Good Timing          & 1 & \textbf{3}  & 0  & 0  & 0  & 0  \\
&Bad Timing           & 7 & \textbf{18} & 1  & 3  & 1  & 3  \\ 
&Plausible            & 2 & 5  & 10 & \textbf{26} & 1  & 3  \\ 
&Implausible          & 5 & 13 & 3  & 8  & 25 & \textbf{64} \\ 
&Unintelligible       & 6 & \textbf{15} & 0  & 0  & 0  & 0  \\ 
&Repetitive           & 3 & 8  & 4  & \textbf{10} & 2  & 5  \\ 
&Vague                & 5 & \textbf{13} & 0  & 0  & 0  & 0  \\ 
\midrule

Vehicle Dynamics & Careful Manoeuvre    & 3 & 8  & 2  & 5  & 4  & \textbf{13} \\ 
&Aggressive Manoeuvre & 1 & 3  & 3  & 8  & 5  & \textbf{13} \\ 
&Vehicle Feature      & 0 & 0  & 3  & \textbf{8}  & 1  & 3  \\ 
   \bottomrule
\end{tabular}
\label{tab:feedback}
}
\end{table*}

\begin{figure*}[h!]
\centering
  \includegraphics[width=\linewidth]{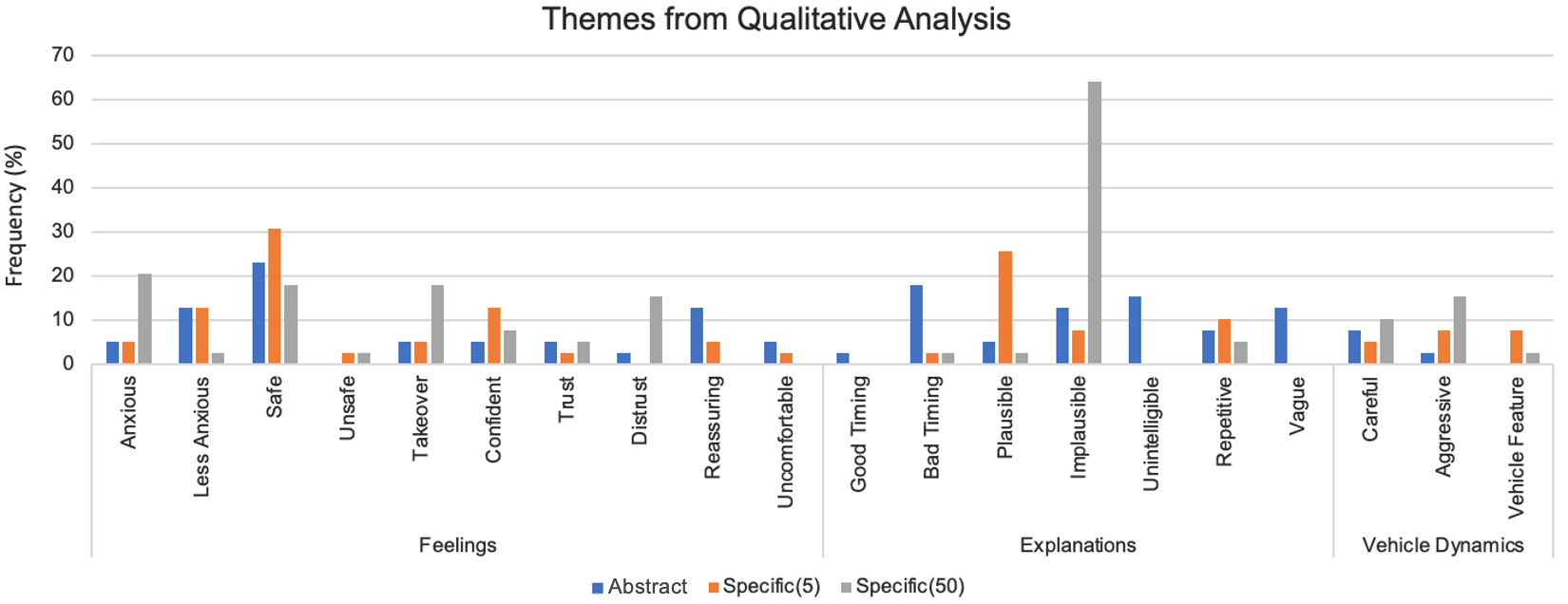}
  \caption[Themes derived from the thematic analysis]{\footnotesize Themes derived from the thematic analysis of the qualitative data from participants. Frequency is expressed in percentage of the total number of responses in each scenario.}~\label{fig:feedback_plot}
\end{figure*}

We obtained qualitative data from the APT questionnaire administered after every scenario. Participants were asked to describe their feelings regarding the explanations they received during the ride. Table~\ref{tab:feedback} and Figure~\ref{fig:feedback_plot} describe the themes obtained from the inductive thematic analysis of the comments. Themes are broadly categorised based on the participants' feelings, their assessment of the explanations, and the vehicle dynamics.

Perceptual errors in the \textit{Specific(50)} scenario evoked negative emotions of anxiety, feeling to takeover navigation control and distrust. CAND1 expressed a feeling of anxiety: `\textit{The explanations made me feel a bit anxious, it says many things that were not right and misleading. I had the urge to look at the buildings and the environment but could not really do that because I wanted to be sure the vehicle is taking the right decision.}'. 
CAND39 expressed the urge to takeover navigation control: `\textit{When the explanations are false, e.g. 'a cyclist is crossing my lane', and it is actually a pedestrian, it made me slightly anxious and likely to want to take over. But nevertheless, I felt safe in the vehicle}'.
CAND5 expressed distrust in the AV: `\textit{anxious as the vehicle did not correctly understand the environment and the types of vehicles around it, which made me trust its judgement less}'. More participants expressed a feeling of safety in the \textit{Specific(5)} scenario:
`\textit{felt safe that the vehicle understood the road and what was going on around us}'.
About the same number of participants expressed a decline in their feeling of anxiety in the \textit{Abstract} and \textit{Specific(5)} scenarios. An example is CAND34's comment about the abstract scenario: `\textit{When the explanations provided are more general, e.g. 'vehicle' instead of 'van' and 'road user' instead of 'cyclist', it feels like the vehicle has a better understanding of the surroundings because it gives a correct explanation, so I felt less anxious and unsafe}'. The abstract explanations might have concealed some errors, in turn, reducing the feeling of anxiety.

There were specific comments about the explanations across the three scenarios. Many participants thought that the explanations in the \textit{Specific(5)} were plausible in that they sounded correct and aligned with what the participants saw. For example: `\textit{Explanations were clear and made sense. Still don't feel some of the reactions were as quick as I might have made them}'---CAND14. There were a good number of comments around the implausible nature of the explanations in the \textit{Specific(50)} scenario. For example, CAND20 said, `\textit{The vehicle this time had difficulty giving the correct reason for stopping/going. Couldn't tell the difference between a pedestrian and a cyclist sometime or thought that traffic lights were off instead of green. I feel that this time I would have wanted more control over the car, particularly at traffic lights as I could determine better if a traffic light was 'working' or not}'.

A couple of candidates thought that the explanations in the \textit{Abstract} scenarios were either too early or late. For example, `\textit{The explanations should have arrived a bit earlier, like a few meters before the vehicle actually stops so that I will know that it is planning to stop. Also, I would be more comfortable if the explanation 'traffic sign' was 'traffic light is red/green'. when referring to a traffic light.}'---CAND19.

Some interesting comments were made about the vehicle's driving style and its interior. For example, CAND31 made a comment about the careful manoeuvre of the vehicle in the \textit{Specific(50)} scenario: `\textit{I was calm throughout the journey. There was no feeling of anxiety as the vehicle did not speed too much to make me feel that way.}'---CAND31. There was a comment relating to aggressive manoeuvre in the \textit{Abstract} scenario: `\textit{Seemed like oncoming vehicles were going to collide with me. It seems to sometime drive on pavements when negotiating corners.}'---CAND35. The rotating steering wheel of the vehicle made some of the participants uncomfortable: `The steering wheel moving abruptly startled me sometimes.'---CAND21 (\textit{Specific(5)} scenario).
Some participants liked the vehicle indicators and the sound they made when indicating the next directions. `\textit{The indicator sound was nice to hear. [...]}'---CAND6 (\textit{Specific(50)} scenario).

\input{discussion}

\section{Conclusion}
In this study, we conducted a rigorous within-subject laboratory investigation ($N=39$) utilising an immersive driving simulator to examine passengers' preferences for explanation specificity levels and the impact of varying perception system errors on perceived safety, and related constructs in autonomous vehicles (AVs). Our findings reveal a link between explanation specificity, perception system errors, and passenger anxiety, contributing to the growing body of literature on trust calibration and transparency in automated vehicles. We observed a significant increase in anxiety levels when fine-grained, intelligible explanations exposed perception system errors. Interestingly, passengers preferred highly detailed (or specific) explanations when AV perception system errors were low, despite the potential for vague explanations to conceal such errors. Hence, as desirable as transparency is, the more we have it beyond certain thresholds, the more anxious we might be. We refer to this phenomenon as the \textit{transparency paradox} in the context of this paper. While our study provides valuable insights, we acknowledge its limitations, particularly, the absence of explanation personalisation, and the inability to select an AV driving style. Future research should address these limitations by investigating the implications of providing passengers with personalisation options, drawing upon theories of user-centred design and adaptive interfaces. Additionally, longitudinal studies and exploration of individual differences could yield valuable information for tailoring AV explanations to diverse user populations. Our findings have significant implications for AV explanation interface design, emphasising the need for adaptive, context-sensitive explanation systems that balance transparency with user comfort, ultimately fostering greater public acceptance and trust in autonomous transportation systems.


\printcredits
\section*{Acknowledgements}
This work was supported by the EPSRC RAILS project
(grant reference: EP/W011344/1) and the EPSRC project
RoboTIPS (grant reference: EP/S005099/1).
\bibliographystyle{cas-model2-names}

\bibliography{bibliography}

\appendix
\input{appendix}

\end{document}

%% file: introduction.tex
\section{Introduction}

The automotive industry is witnessing an increasing level of development in the past decades, from manufacturing manually operated vehicles to manufacturing vehicles with a high level of automation. Despite these technological strides, accidents involving AVs continue to undermine public trust~\cite{model1,teslacrash,ekim7,ekim8,ekim9}.
As highly automated vehicles make high-stake decisions that can significantly affect end-users, the vehicles should explain or justify their decisions to meet set transparency guidelines or regulations, e.g., GDPR Article 12~\cite{gdpr} and the ~\cite{ico}.

Accompanying AVs' driving decisions with natural language explanations is one promising approach for better vehicle transparency~\cite{omeiza2021explanations, ha2020effects, koo2015did}. This transparency, obtained through intelligible explanations, can help to reassure passengers of safety and also assist them in effectively calibrating their trust in an AV~\cite{khastgir2018calibrating}. The specificity level of explanations is, however, an important factor in achieving the aforementioned benefits. In real-world deployments, AVs may not always achieve perfect scene understanding due to the limitations of their perception systems. Depending on the specificity level, explanations might reflect these flaws even when they are inconsequential. Informing operators about these inherent imperfections could enhance safety by helping them recognise when to remotely take control of the vehicle~\cite{kunze2019automation} and it would equally help developers optimise for higher accuracy. While this transparency is helpful for these groups, it remains uncertain whether in-vehicle passengers would prefer high level transparency (if at all useful) that reveals such minor errors (such as mistaking a van for a bus). This is what we refer to the transparency paradox in this paper. It is therefore important to determine the appropriate level of transparency for in-vehicle passengers, mediated through the specificity of the explanations.

Furthermore, as passengers are likely to engage in other activities during their ride, relying solely on visual cues to communicate awareness may be ineffective when passengers' attention is desired. Hence, auditory and vibrotactile feedback\cite{kunze2019conveying} are also necessary to ensure passengers are adequately informed and can respond appropriately in critical situations.

\begin{figure}
     \centering
         \includegraphics[width=6.5cm]{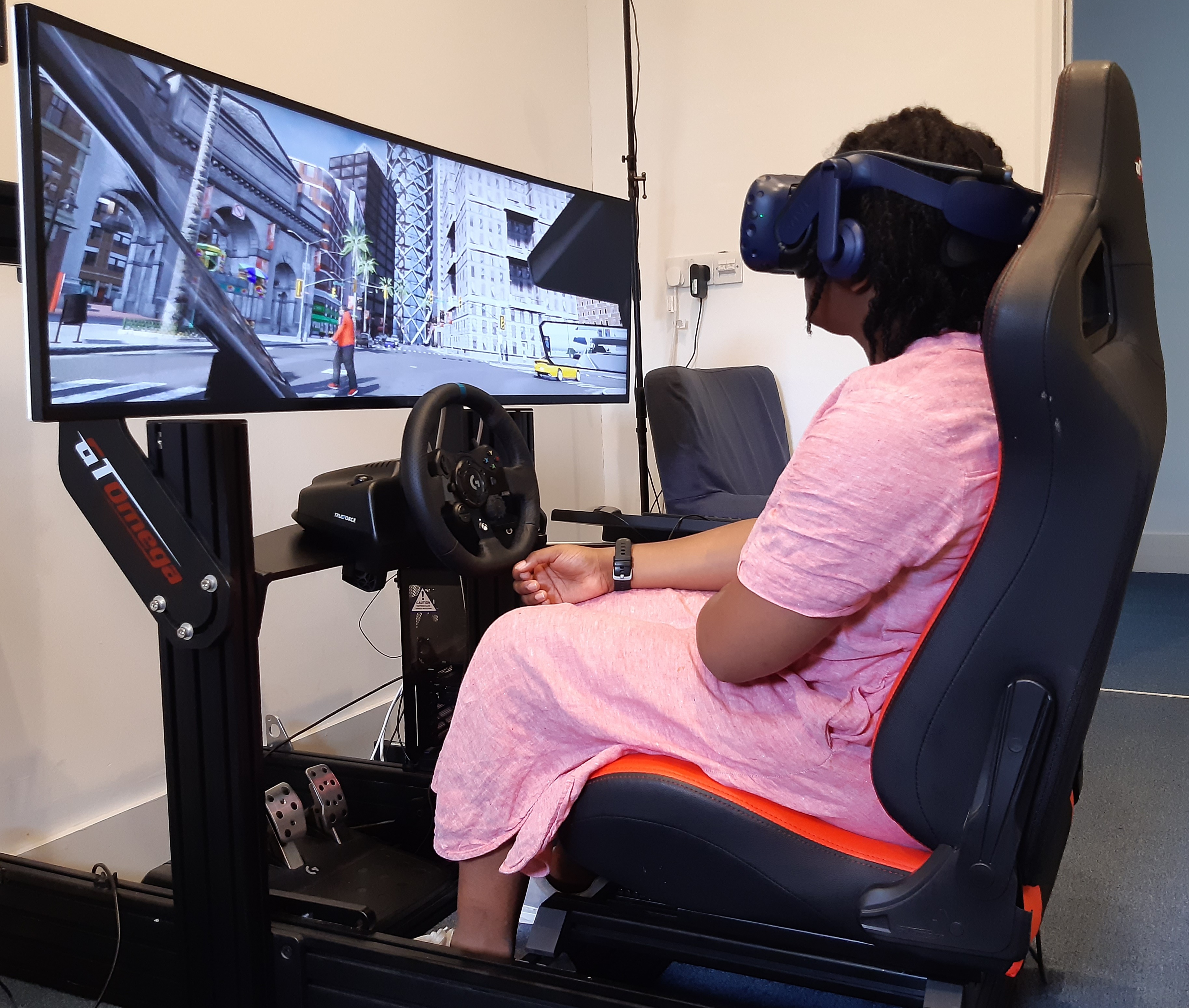}
         \label{fig:participant}
\caption{\footnotesize Driving simulation setup for the study. The setup included a VR headset, steering wheel, brake and acceleration pedals, screen, and arcade seat. The screen shows a pedestrian crossing at a crosswalk.}~\label{fig:setup}
\end{figure}

In this study, we investigate the effects of explanation specificity on AV passengers' perceived safety, feeling of anxiety, and desire to takeover control. We test two levels of explanation specificity: abstract and specific. We use the term \textit{abstract} to describe the provision of vague auditory explanations that conceal details about a driving situation (including perception system errors). In contrast, \textit{specific} refers to the provision of very detailed and fine-grained explanations about a situation. We found non-significant effect of explanation specificity on perceived safety. However, we noticed a significant influence with respect to the amount of perception errors unveiled in the explanations (i.e., between Specific (5) --- Specific (50). Perception of safety dropped in the Specific (50) scenario compared to Specific (5). The feeling of anxiety also significantly increased with the increase in perception errors. There was a significant effect of explanations specificity on takeover feeling (Abstract --- Specific (50)), but no significant difference noticed between Specific (5) and Specific (50). Lastly, no strong correlation was observed between the participants' behavioural cues and the aforementioned factors. 

\subsection{Contribution Statement}

Overall, this research makes three contributions to the fields of explainable autonomous driving and human-machine interaction:

\begin{enumerate}
    \item It sets out a new case study of explanation specificity in the presence of perception systems errors in the autonomous driving context;
    \item It provides an enhanced interpretable technique for generating textual and auditory natural language explanations for AV navigation actions;
    \item  It reveals experimental findings on whether high AV transparency, though critical to other stakeholders, is helpful to AV passengers.
\end{enumerate}

%% file: discussion.tex
\section{General Discussion}
We examined the effects of explanation specificity (\textit{abstract} and \textit{specific}) in AD, while accounting for varying degrees of perception system errors (\textit{low} and \textit{high}). We focused on how this setup would impact passengers' perceived safety and related factors, such as the feeling of anxiety and the thought to takeover control. Our results not only corroborate but also extend previous findings in the field, among others, demonstrating that while intelligible explanations generally create positive experiences for AV users~\cite{omeiza2021towards, ha2020effects, schneider2021explain, m2021calibrating}, this effect is predominantly observed when the AV's perception system errors are low.

\subsection{Psychological Effects}
\paragraph{\textbf{Hypothesis 1.1 - Low transparency yields higher perception of safety}}
Contrary to expectations, participants expressed a greater sense of safety in the \textit{Specific(5)} scenario, indicating a preference for specific explanations in an AV with notably reduced perception system errors. This finding challenges our initial hypothesis and suggests a more nuanced relationship between transparency and perceived safety. It appears that users may prefer more detailed information when the system demonstrates high reliability. 
However, overly detailed explanations were perceived as verbose and repetitive by some participants in the specific scenarios. Therefore, achieving a balanced approach between explanation specificity (or transparency in general) and the cognitive load imposed on passengers is crucial~\cite{poursabzi2021manipulating}. This observation underscores the importance of carefully calibrating the level of transparency to optimise user comprehension and trust.

As observed from Figure~\ref{fig:factors}, AVs characterised by high transparency and high perception system errors generally elicited lower perceptions of safety among passengers. This result aligns with previous work suggesting that transparency can sometimes amplify the negative effects of errors on trust in automated systems~\cite{de2018automation}. In our study, the combination of detailed explanations and frequent errors may have heightened passengers' awareness of the system's limitations, potentially exacerbating their safety concerns. Nonetheless, qualitative responses revealed that a minority of participants reported positive sentiments despite these errors. They praised the vehicle for its ability to detect obstacles and respond appropriately, suggesting that for these individuals, the AV's correct decision-making outweighed concerns about the specific type of obstacle encountered. 

The diverse reactions we observed highlight the complex nature of human-AV interaction and underscore the need for personalised approaches to transparency and explanation design. Adaptive explanation strategies that consider individual differences in information processing and risk perception may be necessary to optimise user experience and safety perceptions in autonomous vehicles.

\paragraph{\textbf{Hypothesis 1.2 - Feeling of Anxiety increases with increasing perception system errors}}
Drivers' anxiety has been noted to increase when utilising AVs \cite{koo2016understanding}. In our study, we anticipated that passengers' feelings of anxiety would correspondingly increase with higher levels of perception system errors in an AV. This hypothesis was supported by a significant difference in anxiety levels observed between the \textit{Specific(5)} and \textit{Specific(50)} scenarios. In the context of AVs, perception system errors represent a critical factor that can directly impact passenger safety and, consequently, their psychological state. As the frequency of errors increases, passengers may experience a heightened sense of uncertainty and loss of control. This might be contributors to the reported anxious feelings.

Given that participants perceived the highest safety in the \textit{Specific(5)} scenario (as seen Hypothesis 1.1), we expected the lowest levels of anxiety in this scenario, assuming a relationship between perceived safety and anxiety, as suggested by prior research~\cite{dillen2020keep}. This inverse relationship between perceived safety and anxiety aligns with the broader psychological concept of risk perception and its impact on emotional states~\cite{stapel2022road, slovic2010feeling}.
Our findings extend beyond the mere confirmation of this relationship, highlighting the nuanced interplay between system transparency, error frequency, and the feeling of anxiety. The significant difference in anxiety levels between the \textit{Specific(5)} and \textit{Specific(50)} scenarios suggests that there may be a threshold of error frequency beyond which anxiety levels sharply increase. This notion is supported by research in risk perception, which indicates that individuals often have a non-linear response to increasing risk levels~\cite{fischhoff1993risk}. Investigating this threshold is an interesting topic for future research.

While Dillen et al., \cite{dillen2020keep} primarily examined how AV driving styles affect driver anxiety and comfort, they noted that certain in-vehicle features, such as a rotating steering wheel, could also influence feelings of anxiety among participants. This experience was reported by participants  (e.g., CAND21) in our study. This observation underscores the multifaceted nature of anxiety in AV contexts, where both system performance and physical design elements play crucial roles.

Furthermore, our results suggest that the relationship between perception system errors and anxiety may be moderated by the level of explanation specificity. In scenarios with high transparency (i.e., specific explanations), the impact of errors on anxiety might be more pronounced, as passengers are more acutely aware of the system's limitations. On the other hand, uncertainty about the workings of algorithms
and perceived lack of control has been noted to cause what researchers term `algorithmic anxiety'~\cite{eiband2019people}. This is also observed from our result in Table~\ref{fig:factors} as the feeling of Anxiety is higher in the \textit{Abstract} scenario compared to \textit{Specific(5)}.

It is worth noting that individual differences in technology acceptance and risk tolerance may also influence the anxiety response to perception system errors~\cite{choi2015investigating}. Future research could explore these individual factors to develop a more comprehensive model of anxiety in AV contexts.

\paragraph{\textbf{Hypothesis 1.3 - Takeover feeling increases with the increase in perception system error}}
Contrary to our hypothesis, the data did not support an increase in takeover feeling with increased perception system errors. While we observed a significant difference between the \textit{Abstract} and \textit{Specific(50)} scenarios, the lack of significant difference between \textit{Specific(5)} and \textit{Specific(50)} scenarios suggests a complex relationship between system errors and user responses. This finding indicates that mere disclosure of perception system errors does not necessarily escalate passengers' desire for control, pointing to a possible threshold effect in how passengers perceive and respond to AV errors. These results challenge the notion that awareness of system imperfections automatically erodes trust in automation~\cite{lee2004trust}, suggesting instead that passengers may be more tolerant of disclosed errors than previously thought, especially with regards to seizing control in autonomous driving.

Our empirical findings contrast with the conceptual analysis presented by~\cite{terken2020toward}, who advocated for shared control between vehicle and user. Our results suggest that users might be more comfortable with full automation than theoretical analyses have predicted, even when aware of system limitations. It is crucial to reconcile these findings with our previous results showing increased anxiety levels as error rates increased. This apparent contradiction suggests a complex interplay between emotional responses and behavioural intentions in AV contexts. While passengers may experience heightened anxiety feelings when aware of higher error rates, this emotional response does not necessarily translate into a stronger desire to take control of the vehicle.
This disconnect between anxiety and takeover feeling could be attributed to factors such as perception of one's own ability to manage the situation, cognitive dissonance between acknowledging risks and maintaining comfort with automation~\cite{aronson1969theory}, or trust in the system's overall capability. 

It is important to note on the other hand that our qualitative results (Table~\ref{tab:feedback}) indicate higher number of comments relating to takeover intents in \textit{Specific (50)} compared to \textit{Specific (5)}. It might be the case that participants might have struggled to calibrate their thoughts or feelings on a 5-point Likert scale. It might also be the case that this difference in frequency of the said comments are statistically insignificant. This would be investigated in future work.

\subsection{Behavioural Cues}
\paragraph{\textbf{Hypothesis 2.1 - Visual signal correlates with anxiety}}
Our study's findings challenge the established link between anxiety and visual distraction proposed by Hepsomali et al.~\cite{hepsomali2017pupillometric}. We found no significant correlation between fixation point divergences and anxiety levels across the different scenarios. This surprising outcome might be attributed to the varied individual priorities in visual attention. For instance, while some participants might have focused on the cityscape, others may have concentrated on areas highlighted by the explanations. The \textit{Specific(50)} scenario exhibited higher divergences in fixation points compared to the \textit{Specific(5)} and \textit{Abstract} scenarios, possibly due to misclassifications in explanations directing attention to incorrect elements. However, the \textit{Specific(5)} and \textit{Abstract} scenarios showed similar fixation effects, suggesting that explanations might have been more effective in these scenarios than in the \textit{Specific(50)} scenario.
Contrary to Dillen et al.,~\cite{dillen2020keep}'s findings on the relationship between eye movement entropy and anxiety, our study did not reveal a significant correlation between saccade differences and anxiety levels. Interestingly, the \textit{Specific(5)} scenario demonstrated the lowest saccade difference, potentially indicating reduced distraction or confusion. This assumption is based on the understanding that saccade velocity reflects the speed of gaze shifts between fixation points. In contrast, the \textit{Abstract} scenario exhibited the highest saccade difference, which could be interpreted as more active visual searching due to the non-specific nature of the explanations failing to effectively guide participants' gaze.

These findings amplifies the existing complexities in the relationship between visual behaviour, anxiety, and the nature of explanations in automated driving scenarios. Further research to better understand the interplay between cognitive states, visual attention, and the effectiveness of different types of explanations in automated driving contexts is in order. Additionally, investigating the potential impact of individual differences in visual processing and attention allocation could provide valuable insights into designing more effective and personalised explanation systems for automated vehicles.

\subsection{Practical Implications}
Our findings challenge the initial assumption that passengers may not desire specific explanations detailing error information from automated vehicles (AVs). Contrary to this presupposition, the study reveals a preference among passengers for specific explanations, particularly when the AV's perception system demonstrates near-perfect accuracy. This insight has significant implications for both manufacturers and regulators in the AV industry.
The observed inverse relationship between perception system errors and passenger anxiety underscores the critical need for highly transparent AVs with exceptional perception and decision-making capabilities. This finding aligns with the broader literature on trust in automation, which emphasises the importance of system reliability and transparency in fostering user acceptance \cite{lee2004trust}. Manufacturers should prioritise the development of robust perception systems that minimise errors, while simultaneously implementing transparent communication mechanisms to convey system status and decision-making processes to passengers. Regulators, in turn, should consider establishing or strengthening existing standards for AV perception accuracy and mandating clear and usable interfaces for conveying this information to passengers.

While our study did not observe direct consequences of misclassification errors on AV actions, it is crucial to recognise the potential ramifications of such errors in more complex, real-world scenarios. The accurate classification of obstacle types is paramount in determining appropriate vehicle responses, particularly when dealing with dynamic obstacles possessing varied manoeuvrability characteristics. This aligns with research efforts on situation awareness in automated systems, which emphasises the importance of accurate environmental perception for effective decision-making~\cite{endsley1995toward}. In intricate traffic scenarios, even minor perceptual inaccuracies could lead to sub-optimal navigation decisions, potentially compromising safety. Therefore, AV developers must strive for highly accurate environmental estimations to ensure appropriate responses to diverse obstacle types, such as bicycles or motorcycles, with very similar attributes but differing capabilities.

The study's findings highlights the link between transparency and accuracy in AV systems. This relationship echoes the concept of `calibrated trust' in automation, where user trust is appropriately aligned with system capabilities~\cite{lee2004trust}. To foster this calibrated trust, AV interfaces should not only provide accurate information but also communicate the system's confidence levels and potential limitations. This approach can help manage passenger expectations and maintain appropriate levels of situational awareness.

While visual feedback from experimental studies provides valuable insights into participants' psychological and behavioural responses, it is useful to simultaneously implement multiple methodologies (e.g., mixed methods) for more accurate conclusions. Complementing observational data with other measurement techniques, such as standardised surveys, physiological measurements, or qualitative interviews, can provide a more comprehensive understanding of user experiences and perceptions. This multi-faceted approach allows for triangulation of data and more confident interpretations of complex human-machine interactions.

Future research should explore the long-term effects of exposure to AV explanations on passenger trust, anxiety, and overall acceptance. Longitudinal studies could reveal how user preferences and responses evolve over time, informing the design of adaptive explanation systems that cater to changing user needs and expectations. Additionally, investigating the impact of cultural differences and individual variability in risk perception on AV explanation preferences could yield insights for developing globally applicable, yet locally tailored, AV communication strategies.

In conclusion, these findings have far-reaching implications for AV design, regulation, and user experience. By prioritising both transparency and accuracy, and adopting a holistic approach to user research, the AV industry can work towards creating systems that not only meet technical performance standards but also address the complex psychological needs of their users, thereby avoiding the transparency paradox and ultimately fostering greater public acceptance and trust in automated transportation technologies.

%% file: appendix.tex
\section{Online Survey Questionnaire}
Note that the same questionnaire was administered in all scenarios.
\begin{figure}
    \begin{center}
    \includegraphics[page=1, scale=0.7]{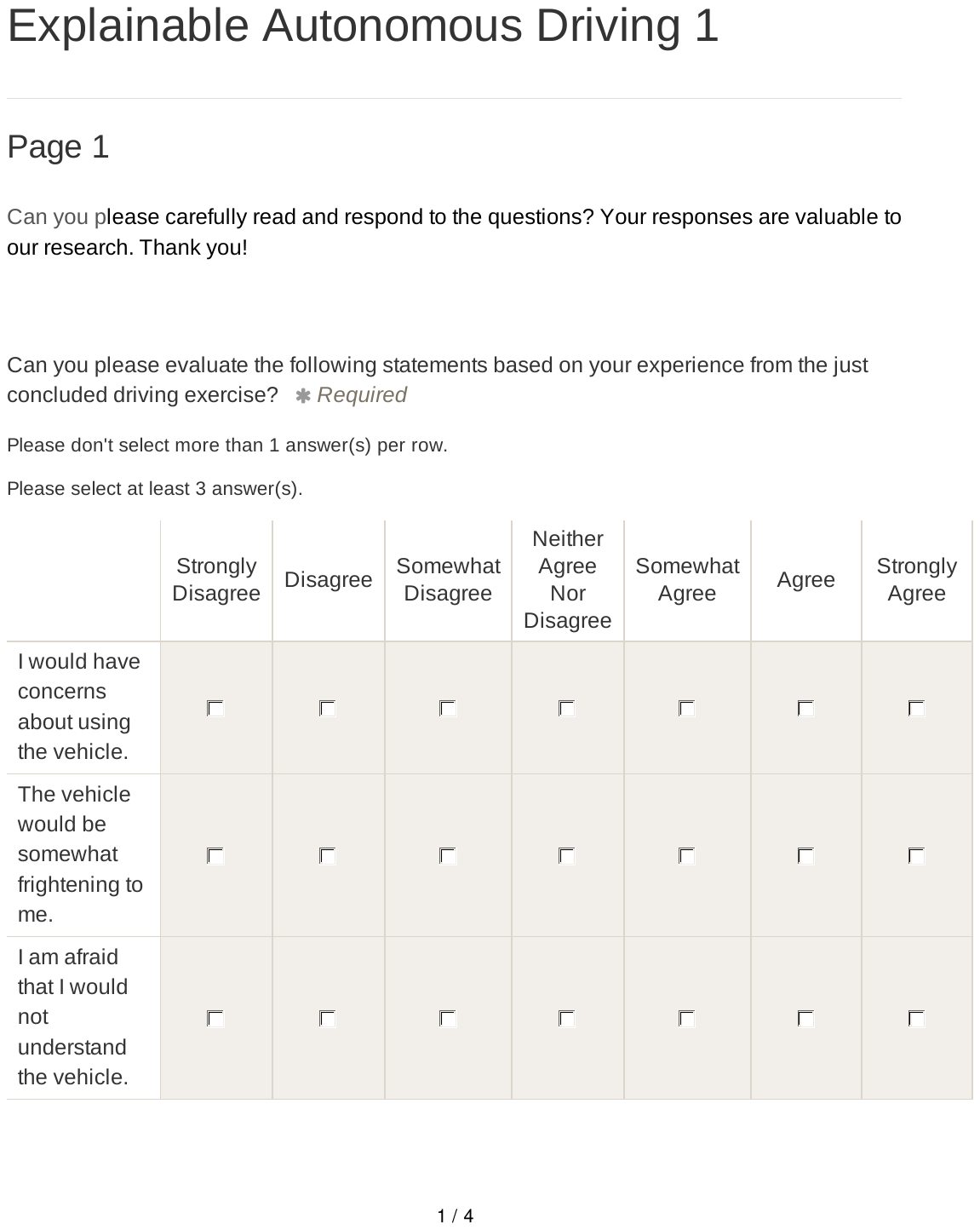}
    \end{center}
\end{figure}

\begin{figure}
    \begin{center}
    \includegraphics[page=2, width=\linewidth]{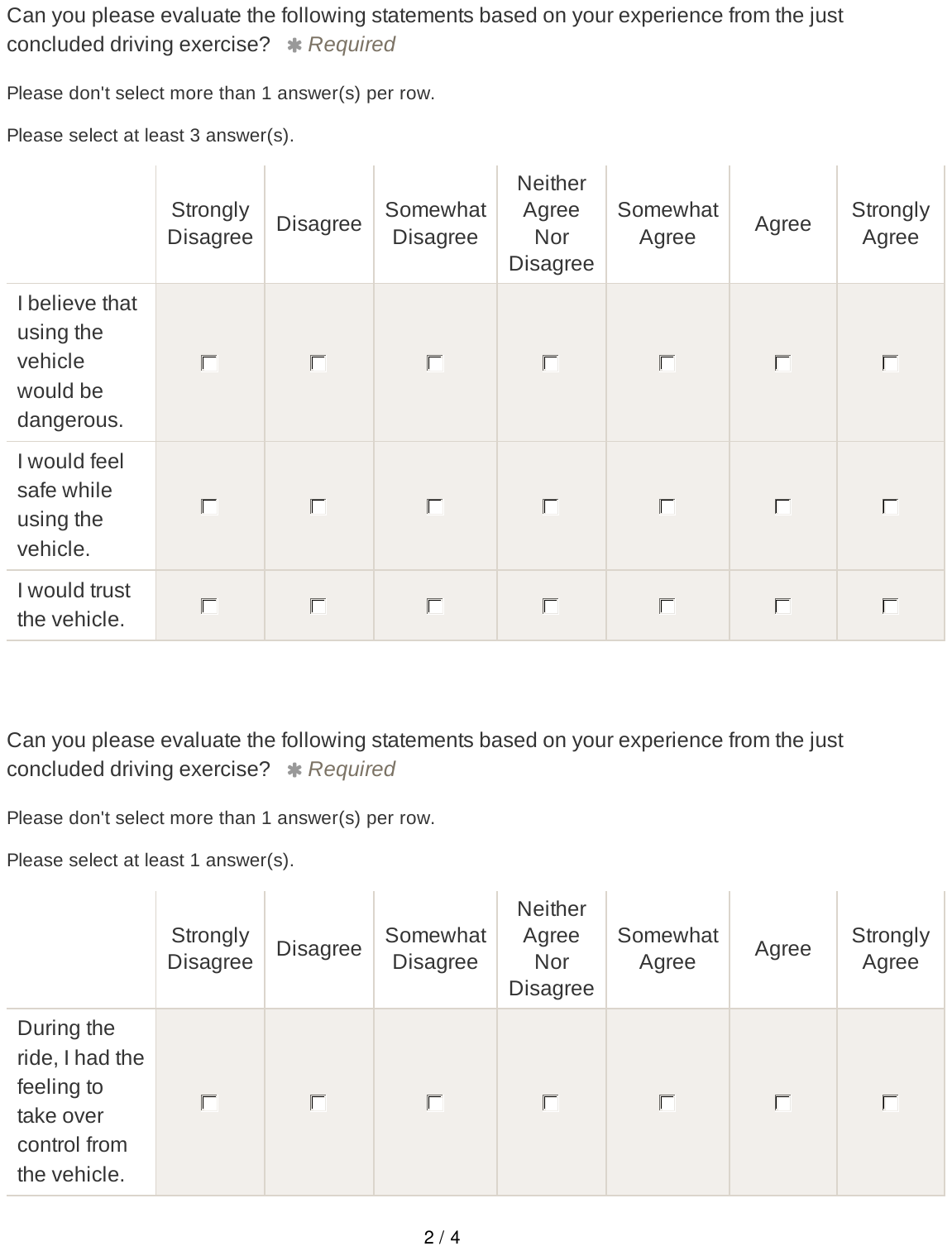}
    \end{center}
\end{figure}

\begin{figure}
    \begin{center}
    \includegraphics[page=3, width=\linewidth]{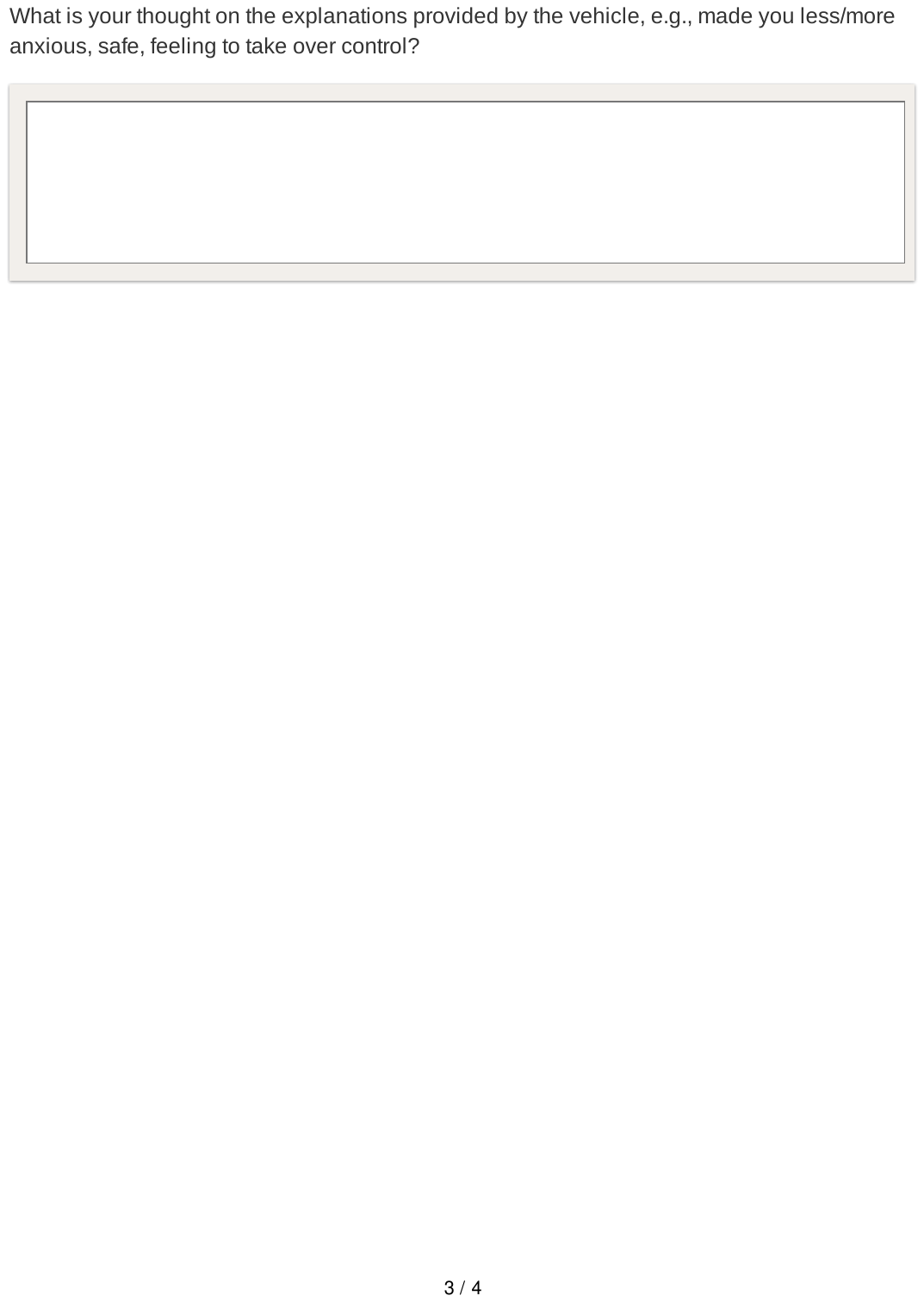}
    \end{center}
\end{figure}

\begin{figure}
    \begin{center}
    \includegraphics[page=4, width=\linewidth]{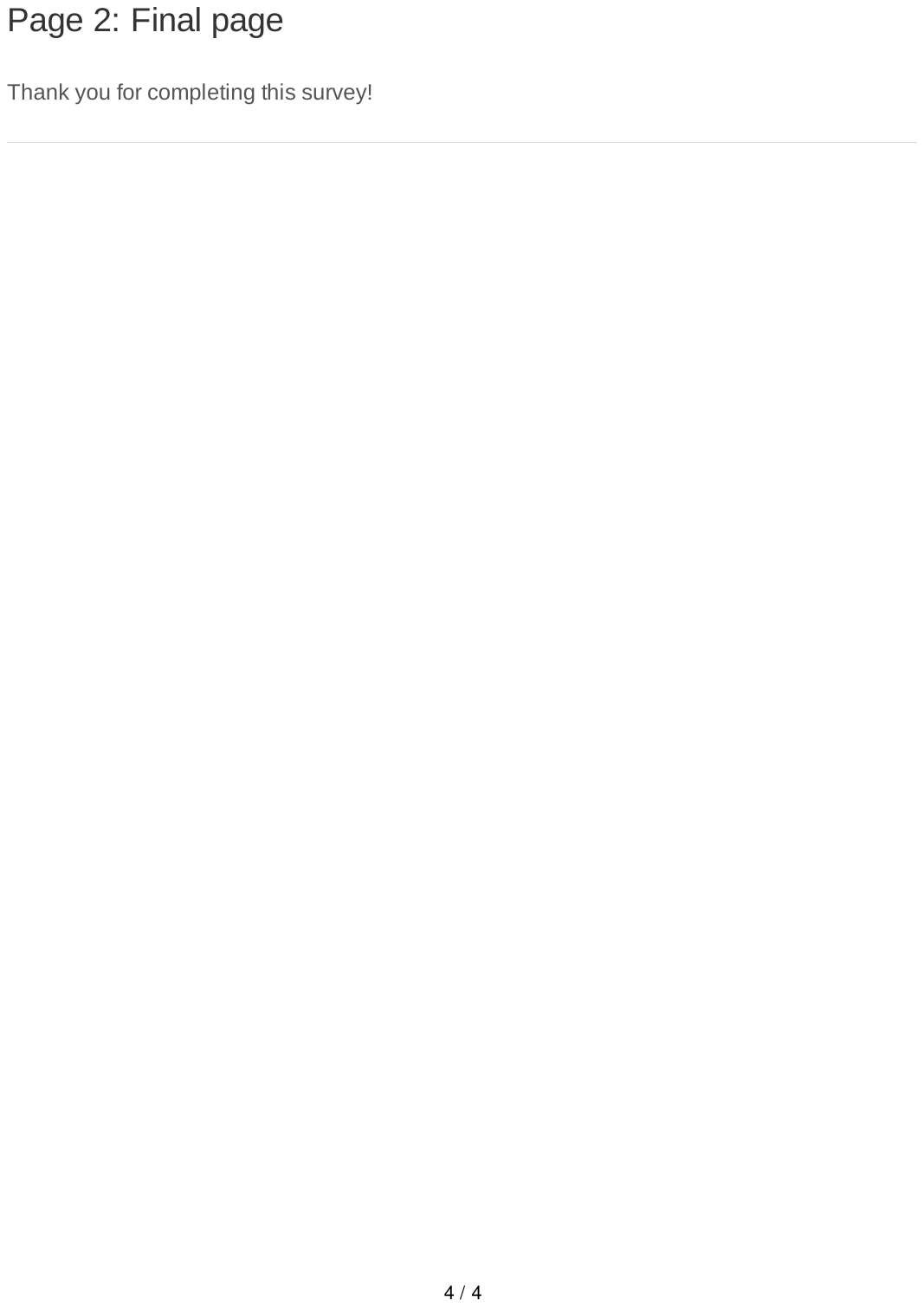}
    \end{center}
\end{figure}